\documentclass[12pt,a4paper]{article}

\usepackage[utf8]{inputenc}
\usepackage[T1]{fontenc}
\usepackage{lmodern}
\usepackage[english]{babel}

\usepackage{geometry}
\usepackage{setspace}
\usepackage{array}
\newcolumntype{C}[1]{>{\centering\arraybackslash}p{#1}}

\usepackage{graphicx}
\usepackage{booktabs}
\usepackage{threeparttable}
\usepackage{array}
\usepackage{longtable}
\usepackage{float}
\usepackage{caption}
\usepackage{tikz}
\usetikzlibrary{arrows.meta, positioning, calc}

\graphicspath{{figures/}}

\usepackage{amsmath, amssymb, amsthm}

\usepackage{xcolor}
\definecolor{linkblue}{RGB}{0,60,120}

\usepackage[round,authoryear]{natbib}
\usepackage[
    colorlinks=true,
    linkcolor=linkblue,
    citecolor=linkblue,
    urlcolor=linkblue
]{hyperref}

\title{Sovereign Stress Avalanches and Network Amplification in Latin America}

\author{Diego Vallarino\thanks{{\fontsize{10}{12}\selectfont
Inter-American Development Bank (IDB),
Washington, DC, United States. Email: \texttt{diegoval@iadb.org}. 
The views expressed in this paper are strictly those of the author and do not necessarily represent the views of the Inter-American Development Bank, its Board of Directors, or the countries they represent.}}
\vspace{0.6cm}
}

\date{April 2026}

\begin{document}
\maketitle

\begin{abstract}
\begin{singlespace}
\noindent
This paper studies sovereign stress avalanches and network amplification in Latin American credit markets using monthly J.P.M-EMBI Global Diversified spreads for eleven sovereigns, 2007–2026. Country stress events are defined as positive log-spread innovations exceeding country-specific volatility thresholds; regional avalanches count the number of stressed countries in each month. The empirical design combines finite-sample power-law diagnostics, threshold robustness, a country-level reshuffling placebo, and rolling correlation, partial-correlation, and minimum-spanning-tree networks. Avalanche sizes are heavy-tailed, with an estimated exponent of 1.77, while spread changes and inter-event times lie in a heavy-tail boundary regime. The placebo shows synchronization far above independent stress timing (p < 0.001). Large avalanches coincide with denser, more spectrally amplifying raw-correlation networks, but not after partial-correlation filtering, indicating common-factor co-movement rather than conditional regional propagation. Network metrics describe contemporaneous stress regimes, not early-warning signals. The results provide a finite-size criticality framework for monitoring sovereign fragility in emerging markets.
\end{singlespace}

\medskip
\noindent\textbf{Keywords:} Self-organized criticality; sovereign spreads;
EMBI; Latin America; financial networks; spectral fragility; contagion;
finite-size criticality.

\noindent\textbf{JEL codes:} G15, F34, C58, E44, F65.
\end{abstract}

\section{Introduction}
\label{sec:intro}

The Latin American sovereign credit market exhibits empirical patterns that
sit uncomfortably with standard macroeconomic intuition. Argentine spreads
multiplied 2.5-fold in a single month after the August 2019 PASO primaries
while Chilean spreads moved within a 50 basis-point envelope; Ecuador
defaulted twice within a decade with peak spreads above $5{,}000$ basis
points, yet Peru held below $300$; in March 2020 every country in the region
experienced a simultaneous large spread blowout. The combination of
country-level idiosyncrasy and synchronized regional stress is anomalous and
calls for analytical tools that capture both phenomena within a single
framework.

The standard sovereign-spread literature explains spread levels through
fiscal and external fundamentals \citep{Edwards1984, EichengreenMody1998,
Longstaff2011, Hilscher2010} and treats contagion as a second-order
correlation phenomenon \citep{ForbesRigobon2002, KaminskyReinhart2000,
KaminskyReinhartVegh2003}. The framework has been productive but has three
limitations relevant to the present paper. First, it does not naturally
explain the heavy-tailed distribution of spread changes
\citep{Mandelbrot1963, Gabaix2009}. Second, it cannot identify whether
observed synchronization reflects genuine network propagation or coincident
reactions to common shocks; correlation-based diagnostics have been
criticized for being driven by heteroscedasticity
\citep{ForbesRigobon2002}. Third, it provides no operational measure of how
close the system is to a regime transition.

A complementary tradition exists in econophysics. Self-organized criticality,
introduced by \citet{BTW1987, BTW1988}, describes systems that endogenously
evolve toward critical states, producing power-law distributions of event
sizes and inter-event times without external parameter tuning. The framework
was introduced into economics by \citet{Bak1993} and \citet{Scheinkman1994}
and developed within the network tradition by \citet{Acemoglu2012}, who
showed that interconnected sectors can endogenously generate heavy-tail
aggregate fluctuations from idiosyncratic shocks. \citet{Sandhu2016}
demonstrated that Ricci curvature on financial networks tracks fragility
prior to market downturns. Despite this conceptual maturity, the SOC
framework has not been systematically applied to sovereign credit markets,
and especially not to Latin America, where eighteen years of EMBI data
covering the global financial crisis, the Eurozone shock, the taper tantrum,
the COVID-19 rout, and the 2022--2025 Fed cycle furnish a natural laboratory.

This paper closes the gap. Five contributions follow. First, the paper
conducts the first empirical SOC-style analysis of LATAM sovereign credit
using monthly EMBI spreads for eleven countries over October 2007 to April
2026. Second, it documents that three independent statistical
objects---avalanche size, absolute spread change, and inter-event
times---all exhibit the heavy-tail boundary regime described by
\citet{StumpfPorter2012}, in which power-law cannot be rejected against
lognormal at conventional significance levels but thin-tailed alternatives
are clearly inappropriate. Third, the paper reformulates the prediction of
which countries are ``critical'': not the high-spread fiscally stressed
economies, but the mid-credibility highly traded ones. Fourth, the paper
introduces a Spectral Fragility Index (SFI) on rolling co-movement networks
as a contemporaneous descriptor of regional fragility regimes. The SFI
discriminates large-avalanche from non-large-avalanche windows in the raw
correlation filter, although this regime difference is not robust to
partial-correlation filtering, which suggests that most of the network
amplification signal observed during stress episodes is driven by
common-factor co-movement rather than by additional conditional regional
propagation. Lead--lag predictive associations at the three-month horizon
are not statistically significant in this sample, so the SFI is presented
as a descriptive surveillance tool rather than as an early-warning
instrument. Fifth, the paper implements a $B=1{,}000$ country-level
reshuffling placebo that formally distinguishes genuine multi-country
synchronization from coincident independent stress. This is the strongest
empirical result of the paper: observed regional synchronization is far
larger than what is achievable under the independence null, with
$p<0.001$ across all four synchronization statistics.

The findings have implications beyond academic SOC. The fact that observed
synchronization is dominated by intermediate-credibility, well-integrated
sovereigns challenges the policy presumption that regional fragility is a
``weak link'' phenomenon driven by peripheral economies. It instead
suggests that fragility in LATAM is an emergent property of capital-market
integration itself: countries such as Brazil, Mexico, Peru, Colombia, and
Panama participate repeatedly in regional stress episodes not because they
are the most fiscally vulnerable, but because they are sufficiently liquid,
investable, and embedded in global emerging-market portfolios to react to
every meaningful shift in international risk appetite.

The paper is also disciplined about what it does not establish. The
heavy-tail evidence is consistent with finite-size criticality but does not
prove a unique asymptotic power law, since lognormal alternatives cannot be
decisively rejected and the number of avalanche events is moderate
($n=40$). The placebo test rules out independent country-level timing as
the source of synchronization but does not separate endogenous regional
propagation from exogenous common-factor responses. The network amplification
result is contemporaneous rather than predictive. These boundaries are
stated explicitly because the contribution of the paper is empirical
discipline, not maximalist claims: LATAM sovereign credit markets display
heavy-tailed, synchronized, threshold-robust stress avalanches whose largest
events coincide with historically meaningful global episodes, and whose
contemporaneous network environment becomes denser and more spectrally
integrated, with the partial-correlation evidence suggesting that this
integration is largely common-factor driven.

The remainder of the paper is organized as follows.
Section~\ref{sec:literature} reviews the literature.
Section~\ref{sec:theory} develops the theoretical framework and states six
formal hypotheses. Section~\ref{sec:data} describes the data and
Section~\ref{sec:methods} the methodology. Section~\ref{sec:results} reports
the empirical results. Section~\ref{sec:discussion} discusses implications
and Section~\ref{sec:conclusion} concludes.

\section{Literature Review}
\label{sec:literature}

This paper lies at the intersection of four literatures: sovereign credit risk
in emerging markets, financial contagion and connectedness, nonlinear dynamics
and self-organized criticality, and network-based measures of systemic
fragility. Its contribution is to bring these strands together in a setting
where sovereign stress is observed not only as a sequence of country-specific
spread movements, but as a finite-size regional propagation process.

The first strand is the empirical literature on sovereign spreads. The early
benchmark is \citet{Edwards1984}, who established the link between sovereign
borrowing costs, default risk, and macroeconomic fundamentals in developing
countries. Subsequent work extended this framework to emerging-market bonds,
showing that spreads reflect both domestic fundamentals and global financial
conditions. \citet{EichengreenMody1998} document the joint role of country
characteristics and market sentiment in emerging-market debt pricing, while
\citet{HilscherNosbusch2010} show that macroeconomic fundamentals, political
risk, and external vulnerabilities jointly shape sovereign risk premia.
\citet{Longstaff2011}, using sovereign CDS data, find that global factors
account for a large share of sovereign credit-risk variation, implying that
sovereign spreads are not purely national objects but also reflect common
financial conditions. This finding is especially relevant for Latin America,
where the EMBI market is shaped by global risk appetite, benchmark
rebalancing, and regional portfolio allocation.

A related literature emphasizes the historical recurrence of sovereign debt
crises and the nonlinear nature of sovereign-risk regimes. \citet{ReinhartRogoff2009}
document long-run cycles of default, financial repression, and debt overhang,
while \citet{ReinhartTrebesch2016} show that sovereign crises are embedded in
global waves of capital flows, commodity prices, and international liquidity.
Theoretical models of sovereign debt have also emphasized coordination and
multiple equilibria. \citet{Calvo1988} shows how self-fulfilling expectations
can generate debt crises even when fundamentals do not uniquely determine the
outcome, while \citet{ColeKehoe2000} develop a model in which rollover crises
arise from coordination failures among creditors. These contributions suggest
that sovereign stress may involve threshold behavior and regime shifts, but
they do not directly measure the cross-country synchronization of stress
events. The present paper complements this literature by treating synchronized
spread jumps as empirical avalanches and by studying their distribution and
network environment.

The second strand is the literature on contagion, spillovers, and financial
connectedness. \citet{ForbesRigobon2002} provide the canonical critique of
correlation-based contagion tests, showing that correlations mechanically rise
during high-volatility periods and may therefore overstate contagion. This
critique is central to the design of the present paper, which reports raw
correlation networks but gives greater interpretive weight to
partial-correlation and filtered network structures. \citet{KaminskyReinhart2000}
and \citet{KaminskyReinhartVegh2003} identify trade, financial integration,
common lenders, and investor behavior as channels through which crises can
spread across emerging markets. \citet{BekaertHarvey2014} and
\citet{Bekaert2014ContagionEquity} further show that political and
institutional shocks can generate cross-border asset-price responses that are
not reducible to direct fundamental exposure. In the connectedness tradition,
\citet{DieboldYilmaz2014} introduce variance-decomposition-based measures of
systemic connectedness, providing a framework for quantifying directional
spillovers across financial markets. The present paper differs from this
literature by focusing not on average spillover intensity, but on the size
distribution of synchronized sovereign-stress events and on whether network
amplification rises around large avalanches.

The third strand concerns the network origins of aggregate fluctuations and
systemic risk. \citet{Acemoglu2012} show that idiosyncratic shocks can
generate aggregate fluctuations when the production network is sufficiently
asymmetric, thereby overturning the standard diversification logic. In a
related financial setting, \citet{Acemoglu2015} show that interbank networks
can be either shock absorbers or shock amplifiers depending on the size of
shocks and the topology of the system. \citet{HaldaneMay2011} argue that
financial complexity can increase systemic fragility when dense connectivity
facilitates contagion rather than diversification. \citet{GlassermanYoung2016}
clarify the conditions under which financial networks amplify or dampen
losses, emphasizing the role of leverage, concentration, and propagation
structure. These papers motivate the spectral approach used here. In a linear
propagation system, the spectral radius of the weighted adjacency matrix
captures proximity to the stability boundary; as the spectral radius rises,
the system becomes more capable of amplifying shocks. The Spectral Fragility
Index used in this paper is a reduced-form operationalization of this idea in
the context of sovereign-spread co-movement networks.

The fourth strand is the literature on nonlinear dynamics, heavy tails, and
self-organized criticality. \citet{Mandelbrot1963} first documented the
empirical inadequacy of Gaussian models for speculative price changes.
\citet{BTW1987} and \citet{BTW1988} introduced self-organized criticality as a
mechanism through which locally interacting systems endogenously approach a
critical state and generate power-law event-size distributions. \citet{Bak1996}
provides the canonical exposition of sandpile dynamics and avalanche
distributions. In economics, \citet{Bak1993} and
\citet{ScheinkmanWoodford1994} argue that sectoral interactions can generate
aggregate fluctuations with critical properties, while \citet{LuxMarchesi1999}
show that agent-based financial markets can produce scaling laws through
endogenous interaction among heterogeneous traders. \citet{Gabaix2003} and
\citet{Gabaix2009} review the prevalence of power laws in economic and
financial phenomena, and \citet{Sornette2003} develops the concept of
critical events and ``dragon-kings'' as extreme outcomes generated by
endogenous amplification mechanisms.

The empirical challenge in this literature is that finite samples often do
not allow clean discrimination between power-law and lognormal alternatives.
\citet{Clauset2009} provide the standard maximum-likelihood framework for
estimating power-law tails and selecting the lower cutoff, while
\citet{Vuong1989} provides the non-nested likelihood-ratio logic used to
compare power-law and lognormal specifications. \citet{StumpfPorter2012}
stress that many empirical claims about power laws are overstated because
several heavy-tailed alternatives can fit the same data. This warning is
directly relevant here. The present paper therefore avoids claiming that
Latin American EMBI spreads obey a pure asymptotic power law. Instead, it
tests whether avalanche sizes, absolute spread changes, and inter-event times
lie in a heavy-tail boundary regime consistent with finite-size criticality.
The object of interest is not the isolated distribution of $|\Delta s|$ alone,
but the joint evidence of heavy-tailed avalanches, non-random synchronization,
and network amplification.

A further methodological strand concerns filtered financial networks and
geometric measures of fragility. \citet{Mantegna1999} introduced the minimum
spanning tree as a filtering device for extracting the hierarchical structure
of financial correlations, and \citet{Tumminello2005} extended the logic of
information filtering in financial networks. Ricci curvature has more
recently been used to measure the fragility of networked financial systems.
\citet{Forman2003} introduced a combinatorial curvature for weighted cell
complexes, while \citet{Ollivier2009} developed a probabilistic notion of
Ricci curvature based on optimal transport between neighborhood measures.
\citet{Sandhu2016} apply Ricci curvature to financial networks and show that
curvature-based measures can track market fragility prior to crises. In the
present paper, curvature is treated as an auxiliary measure of geometric
integration, while the main amplification results are based on density,
node-strength concentration, spectral radius, and the Spectral Fragility
Index. This distinction is important because sovereign-spread networks during
global stress often become dense and geometrically integrated, so positive
curvature may reflect redundant co-movement rather than bottleneck fragility.

The paper is also connected to recent work on nonlinear dynamic graphs in
economic systems. \citet{vallarino2026survival} studies stochastic network
survival models in which propagation on economic graphs shapes time-to-event
outcomes. \citet{vallarino2026identification} examines identification and
inference in nonlinear dynamic network models, emphasizing dependence,
spectral structure, and empirical interpretation. \citet{vallarino2026sandpile}
develops a theoretical framework for sandpile economics, linking threshold
accumulation, endogenous stress release, and heavy-tailed event distributions.
The present paper complements this work by moving from theoretical and
methodological formulations to an empirical sovereign-credit setting, where
stress avalanches are measured directly from EMBI data and related to
time-varying network amplification.

The gap addressed by this paper is therefore precise. Sovereign-spread models
explain why countries become risky; contagion models explain why spreads
co-move; network models explain how shocks propagate; and criticality models
explain why event sizes may be heavy-tailed. What is missing is an empirical
framework that treats sovereign stress as a finite-size regional avalanche
process and tests whether large synchronized events are statistically
non-trivial and dynamically related to network amplification. This paper
provides that framework for Latin American sovereign credit markets.

\section{Theoretical Framework and Hypotheses}
\label{sec:theory}

\subsection{A finite-size sovereign-stress system}

Consider a finite regional system of $n=11$ Latin American sovereigns indexed
by $c \in \{1,\ldots,n\}$ and observed at monthly frequency. Let
$s_{c,t}>0$ denote the stripped EMBI Global Diversified spread of sovereign
$c$ at month $t$, and define the monthly log-spread innovation as
\begin{equation}
x_{c,t} \equiv \Delta \log s_{c,t}
= \log s_{c,t}-\log s_{c,t-1}.
\label{eq:x_def}
\end{equation}
The empirical object of interest is not only the marginal distribution of
$x_{c,t}$, but the cross-sectional synchronization of large positive
realizations of $x_{c,t}$ across countries.

To motivate the empirical construction, suppose that each sovereign has an
unobserved stress stock $z_{c,t}$ that evolves according to a threshold-based
network propagation process:
\begin{equation}
z_{c,t+1}
=
(1-\delta_c)z_{c,t}
+
\beta_c g_t
+
\gamma_c f_{c,t}
+
\sum_{j\neq c} a_{jc,t}\,
\mathbf{1}\!\left(z_{j,t}>z_j^{*}\right)
+
\varepsilon_{c,t+1},
\label{eq:latent_stress}
\end{equation}
where $g_t$ is a common global financial factor, $f_{c,t}$ denotes
country-specific fiscal, political, or external vulnerability news,
$a_{jc,t}$ captures time-varying regional transmission from sovereign $j$ to
sovereign $c$, and $\delta_c\in(0,1)$ governs stress dissipation. A stress
event occurs when the latent stock crosses a sovereign-specific threshold
$z_c^{*}$. The observed spread innovation $x_{c,t}$ is interpreted as a noisy
market manifestation of changes in the latent stress stock.

The empirical stress indicator is therefore defined as a country-normalized
positive tail event:
\begin{equation}
I_{c,t}(\sigma)
=
\mathbf{1}
\left[
x_{c,t}>\sigma \hat{\sigma}_c
\right],
\qquad
\hat{\sigma}_c
=
\operatorname{sd}(x_{c,t}),
\label{eq:stress_def}
\end{equation}
where $\sigma>0$ is a multiplicative threshold and the baseline specification
uses $\sigma=1.5$. This definition identifies large positive spread innovations
relative to each country's own historical volatility, rather than imposing a
common basis-point threshold on sovereigns with very different spread levels
and volatilities.

A sovereign-stress avalanche is the cross-sectional number of countries in
stress in month $t$:
\begin{equation}
S_t(\sigma)
=
\sum_{c=1}^{n} I_{c,t}(\sigma).
\label{eq:avalanche_def}
\end{equation}
The process $\{S_t(\sigma)\}$ is a finite-size regional stress-size process.
Large values of $S_t$ indicate synchronized sovereign stress, while small
values correspond to isolated country-level stress events.

\subsection{Finite-size criticality and avalanche predictions}

In the canonical self-organized criticality framework, locally interacting
systems endogenously approach a critical state in which small perturbations
can generate events of very different sizes \citep{BTW1987,BTW1988,Bak1996}.
For an infinite system, the central prediction is a power-law distribution of
event sizes. In a finite empirical sovereign system, however, the support of
$S_t$ is mechanically bounded by $n=11$. The appropriate empirical claim is
therefore not that the system satisfies an asymptotic power law, but that it
exhibits finite-size signatures consistent with critical propagation.

The main SOC-consistent empirical prediction is that avalanche sizes are
heavy-tailed:
\begin{equation}
\Pr(S_t \geq s)
\sim
s^{-(\alpha-1)}
\quad \text{for } s\geq s_{\min},
\label{eq:ccdf_powerlaw}
\end{equation}
with an exponent in a range compatible with finite-size critical behavior.
The second prediction is that stress recurrence is irregular rather than
Poisson-like, implying heavy-tailed inter-event times. Let
$\tau_{c,k}$ denote the waiting time between consecutive stress events for
country $c$. Then a critical propagation process should generate a dispersed
distribution of $\tau_{c,k}$, reflecting clustered stress episodes separated
by calm periods. The third prediction is that the underlying spread innovations
also display heavy-tailed magnitudes, although $|\Delta s_{c,t}|$ is not the
primary SOC object. Individual spread changes may be shaped by jumps,
liquidity shocks, and heterogeneous volatility, whereas the avalanche-size
distribution captures regional synchronization.

Following \citet{Clauset2009}, power-law exponents are estimated by maximum
likelihood with the lower cutoff selected by minimizing the Kolmogorov--Smirnov
distance. Following \citet{Vuong1989}, power-law fits are compared with
lognormal alternatives through non-nested likelihood-ratio tests. The
interpretation follows \citet{StumpfPorter2012}: in finite samples, failure to
reject the lognormal alternative does not invalidate the heavy-tail claim.
Rather, the relevant question is whether the data lie in a heavy-tail boundary
regime in which Gaussian or exponential alternatives are clearly inappropriate
and the power-law/lognormal distinction is empirically non-degenerate.

\subsection{Dynamic network amplification}

The latent propagation process in equation \eqref{eq:latent_stress} implies
that stress transmission depends on the time-varying network structure
$A_t=[a_{ij,t}]$. Empirically, the network is not directly observed. I
therefore approximate the sovereign-stress transmission structure using
rolling co-movement networks of EMBI spread innovations.

For each 24-month rolling window, three filtered networks are constructed on
the common set of sovereigns $V=\{1,\ldots,n\}$. The first is a correlation
network, where an undirected edge is included when
$|\rho_{ij,t}|>\rho^{*}$, with $\rho^{*}=0.4$. The second is a
partial-correlation network. Let $\Sigma_t$ denote the covariance matrix of
spread innovations in the rolling window and let
$\Omega_t=\Sigma_t^{-1}$ denote the corresponding precision matrix. The
partial correlation between sovereigns $i$ and $j$ is
\begin{equation}
\rho^{\mathrm{part}}_{ij,t}
=
-\frac{\Omega_{ij,t}}
{\sqrt{\Omega_{ii,t}\Omega_{jj,t}}},
\label{eq:partial_corr}
\end{equation}
and an edge is included when
$|\rho^{\mathrm{part}}_{ij,t}|>\rho^{*}$. The third network is the minimum
spanning tree computed from correlation distances
$d_{ij,t}=1-|\rho_{ij,t}|$. The correlation network captures total
co-movement, the partial-correlation network filters common linear dependence,
and the minimum spanning tree provides a sparse topology with a fixed number
of edges.

For each network, let $W_t=[w_{ij,t}]$ denote the weighted adjacency matrix.
The first dynamic amplification measure is the normalized spectral radius:
\begin{equation}
\rho^{\infty}_t
=
\lambda_{\max}
\left(
\frac{W_t}{\|W_t\|_{\infty}}
\right),
\qquad
\rho^{F}_t
=
\lambda_{\max}
\left(
\frac{W_t}{\|W_t\|_{F}}
\right),
\label{eq:spectral_radius}
\end{equation}
where $\|W_t\|_{\infty}$ is the maximum row-sum norm and $\|W_t\|_{F}$ is the
Frobenius norm. These normalizations avoid the mechanical unit eigenvalue
that would arise from row-stochastic normalization. The Spectral Fragility
Index is defined as
\begin{equation}
\mathrm{SFI}_t
=
\frac{\rho_t}{1-\rho_t},
\label{eq:sfi}
\end{equation}
where $\rho_t$ denotes either normalized spectral radius. In a linear
propagation system of the form
\begin{equation}
y_t = W_t y_{t-1}+\eta_t,
\label{eq:linear_propagation}
\end{equation}
the cumulative response to shocks is governed by the Neumann series
$(I-W_t)^{-1}$, which becomes more amplifying as the spectral radius approaches
the stability boundary. The SFI is therefore interpreted as a reduced-form
indicator of the network's proximity to high-amplification propagation.

\subsection{Network geometry}

In addition to spectral amplification, I compute geometric summaries of the
sovereign-spread network. The Forman--Ricci curvature of edge $(u,v)$ is
defined as
\begin{equation}
\kappa_{uv}^{F}
=
w_{uv}
\left[
\frac{w_u+w_v}{w_{uv}}
-
\sum_{\substack{e_u\sim u\\ e_u\neq uv}}
\frac{w_{uv}}{\sqrt{w_{uv}w_{e_u}}}
-
\sum_{\substack{e_v\sim v\\ e_v\neq uv}}
\frac{w_{uv}}{\sqrt{w_{uv}w_{e_v}}}
\right],
\label{eq:forman}
\end{equation}
following \citet{Forman2003}, where
$w_u=\sum_{k\sim u} w_{uk}$ is node strength. I also compute a normalized
Ollivier-style curvature based on optimal transport between neighborhood
measures. The canonical Ollivier--Ricci curvature on edge $(u,v)$ is
\begin{equation}
\kappa^{O}_{uv}
=
1-
\frac{W_1(\mu_u,\mu_v)}{d(u,v)},
\label{eq:ollivier}
\end{equation}
where $W_1$ is the 1-Wasserstein distance between the probability measures
induced on the neighborhoods of $u$ and $v$ \citep{Ollivier2009}. In the
empirical implementation, distances are normalized to avoid spurious curvature
explosions in highly correlated networks. Curvature is therefore interpreted
as an auxiliary measure of geometric integration rather than as the primary
measure of fragility. The main fragility indicators are spectral radius,
SFI, density, and node-strength concentration.

\subsection{Hypotheses}

The theoretical framework implies six testable hypotheses. They are stated in
terms of empirical objects that are directly measured in the data and tested
in Section \ref{sec:results}.

\begin{description}

\item[\textbf{H1. Avalanche-size heavy tails.}]
The distribution of sovereign-stress avalanche sizes is heavy-tailed and lies
in a finite-size criticality regime. Operationally, the estimated tail exponent
for $S_t(\sigma)$ should lie in a plausible critical range, and power-law
behavior should not be decisively rejected in favor of lognormal alternatives.

\item[\textbf{H2. Non-random synchronization.}]
Large sovereign-stress avalanches are not mechanically generated by
country-specific stress frequencies. Operationally, a reshuffling placebo that
preserves each country's number of stress events but randomizes their timing
should generate substantially smaller maximum avalanche sizes and lower shares
of large avalanches than the observed data.

\item[\textbf{H3. Threshold robustness.}]
The avalanche result should not be an artifact of the baseline threshold.
Operationally, the estimated tail exponent and maximum avalanche size should
remain within a stable range for stress thresholds between
$\sigma=1.25$ and $\sigma=2.00$, with additional diagnostics reported up to
$\sigma=2.50$.

\item[\textbf{H4. Heterogeneous country participation.}]
Country participation in stress avalanches is heterogeneous. Operationally,
stress-event counts should reject uniform participation across countries, and
the ranking of participation should not be mechanically ordered by average
spread level.

\item[\textbf{H5. Network amplification during large avalanches.}]
Large avalanche months are associated with more amplifying network states.
Operationally, rolling network windows overlapping large stress episodes
should display higher density, higher normalized spectral radius, higher SFI,
or lower concentration than non-stress windows.

\item[\textbf{H6. Predictive network association.}]
We test whether network amplification contains lead--lag
predictive content for near-future regional stress.
Operationally, we examine whether current network metrics
in any of the three filters are associated with the maximum
avalanche size observed over the subsequent three months.
A finding of no significant association would indicate that
the network metrics constructed here are descriptors of
contemporaneous stress regimes rather than predictive
instruments at the three-month horizon.

\end{description}

\section{Data}
\label{sec:data}

The empirical analysis uses monthly J.P.~Morgan EMBI Global Diversified
stripped spreads for eleven Latin American sovereigns: Argentina (ARG),
Brazil (BRA), Chile (CHL), Colombia (COL), Dominican Republic (DOM), Ecuador
(ECU), Mexico (MEX), Panama (PAN), Peru (PER), El Salvador (SLV), and Uruguay
(URY). The sample runs from October 2007 to April 2026, yielding a balanced
monthly panel of 223 observations per country and 2{,}453 country-month
observations in total. Spreads are expressed in basis points and are used as
market-based measures of sovereign credit risk. The use of stripped spreads
focuses the analysis on the sovereign credit-risk component embedded in EMBI
bond prices, abstracting from collateralized components and other mechanical
features of the underlying instruments.

The country coverage is chosen to maximize cross-sectional consistency within
Latin America while preserving a long monthly time dimension. The resulting
panel covers a heterogeneous set of sovereigns. Argentina, Ecuador, and El
Salvador display substantially higher average spreads and larger right tails,
whereas Chile, Peru, Uruguay, Panama, Brazil, Colombia, and Mexico trade at
lower average spreads over most of the sample. This heterogeneity is useful
for the purposes of the paper because it allows the empirical analysis to
distinguish between two different notions of fragility: high average sovereign
risk and frequent participation in synchronized regional stress episodes.

Table \ref{tab:data_coverage} reports the country-level coverage and
descriptive statistics. All eleven series are observed without gaps over the
same sample window. Argentina and Ecuador exhibit the largest maximum spreads,
with peaks of 3{,}803 and 5{,}129 basis points respectively, both associated
with their 2020 sovereign credit events. Chile is the most stable sovereign in
the panel, ranging between 87 and 392 basis points. These differences in
levels and tail dispersion motivate the country-normalized event definition
used in the next section: stress is measured relative to each sovereign's own
historical spread-change volatility rather than through a common basis-point
threshold.

\begin{table}[!htbp]
\centering
\footnotesize
\begin{threeparttable}
\caption{Coverage and descriptive statistics of monthly EMBI Global Diversified stripped spreads, by country (basis points).}
\label{tab:data_coverage}
\begin{tabular}{llllrrrrrr}
\toprule
Code & Country & Start & End & $N$ & Mean & Median & SD & Min & Max \\
\midrule
ARG & Argentina & 2007-10 & 2026-04 & 223 & 1,094 & 817 & 669 & 312 & 3,803 \\
BRA & Brazil & 2007-10 & 2026-04 & 223 & 260 & 241 & 75 & 140 & 548 \\
CHL & Chile & 2007-10 & 2026-04 & 223 & 156 & 144 & 47 & 87 & 392 \\
COL & Colombia & 2007-10 & 2026-04 & 223 & 248 & 223 & 89 & 112 & 575 \\
ECU & Ecuador & 2007-10 & 2026-04 & 223 & 1,077 & 824 & 775 & 361 & 5,129 \\
MEX & Mexico & 2007-10 & 2026-04 & 223 & 282 & 271 & 94 & 121 & 656 \\
PAN & Panama & 2007-10 & 2026-04 & 223 & 203 & 187 & 70 & 102 & 582 \\
PER & Peru & 2007-10 & 2026-04 & 223 & 182 & 168 & 60 & 107 & 530 \\
DOM & Dom. Rep. & 2007-10 & 2026-04 & 223 & 403 & 362 & 210 & 164 & 1,705 \\
URY & Uruguay & 2007-10 & 2026-04 & 223 & 198 & 181 & 111 & 64 & 834 \\
SLV & El Salvador & 2007-10 & 2026-04 & 223 & 610 & 459 & 430 & 158 & 2,704 \\
\bottomrule
\end{tabular}
\begin{tablenotes}\footnotesize
\item Source: J.P.~Morgan EMBI Global Diversified (monthly stripped spreads).
\item All series cover 2007:M10 to 2026:M4 ($N=223$ months) without gaps.
\item Spreads in basis points. ARG and ECU exhibit extreme right tails reflecting the 2020 sovereign credit events.
\end{tablenotes}
\end{threeparttable}
\end{table}

The sample contains several major episodes of global and regional financial
stress: the global financial crisis and post-Lehman emerging-market repricing
(2008--2009), the European sovereign crisis and associated global risk-off
episodes (2010--2012), the taper tantrum and emerging-market repricing
(2013--2014), the commodity-price downturn and emerging-market sell-off
(2015--2016), the Argentina currency and sovereign-risk crisis (2018--2019),
the COVID-19 global financial shock (2020), and the Federal Reserve tightening
cycle and post-pandemic repricing of emerging-market risk (2022--2025). The
sample therefore spans both global common shocks and country-specific
sovereign events, which is essential for separating ordinary co-movement from
synchronized regional stress.

\begin{figure}[H]
\centering
\includegraphics[width=\linewidth]{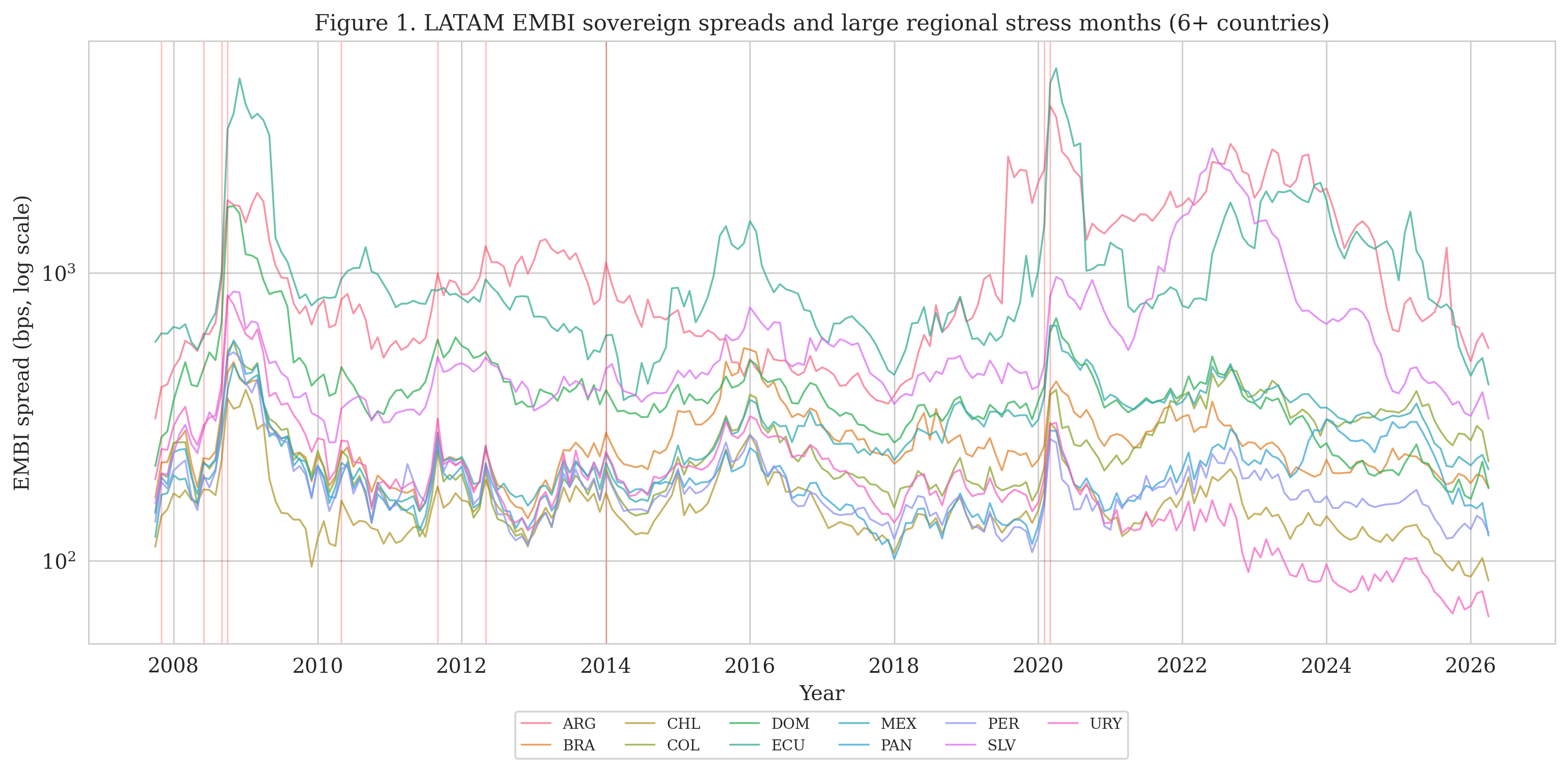}
\caption{LATAM EMBI Global Diversified stripped sovereign spreads,
2007:M10--2026:M4. Spreads are monthly observations in basis points and are
shown on a logarithmic scale. Vertical red lines identify large regional
stress months, defined as months in which at least six sovereigns
simultaneously exceed their country-specific stress threshold
($S_t\geq 6$).}
\label{fig:spreads}
\end{figure}

Figure \ref{fig:spreads} displays the eleven EMBI spread series on a
logarithmic scale. Vertical red lines mark months in which at least six
countries simultaneously exceed their country-specific stress thresholds. The
figure highlights the empirical object studied in the paper: large regional
stress months are not defined by the level of any single sovereign spread, but
by the cross-sectional synchronization of large positive spread changes across
the region. This distinction is central to the avalanche approach developed
below.

The empirical analysis uses the spread panel in three ways. First, monthly
log-spread changes are used to define country-level stress events. Second,
the number of simultaneous stress events in each month defines the
sovereign-stress avalanche size. Third, rolling windows of spread changes are
used to construct time-varying co-movement networks. The same underlying EMBI
panel therefore supports the three empirical layers of the paper: tail
diagnostics, synchronization tests, and network-amplification analysis.

\section{Methodology}
\label{sec:methods}

This section describes the empirical procedure used to identify sovereign-stress
avalanches, estimate tail behavior, construct filtered spread networks, and
evaluate whether observed synchronization exceeds what would be expected under
country-level stress frequencies alone. All tables are stored in the
\texttt{tables/} folder and all figures in the \texttt{figures/} folder.

\subsection{Avalanche identification}
\label{subsec:avalanche_identification}

Let $s_{c,t}$ denote the EMBI stripped spread of sovereign $c$ in month $t$,
measured in basis points. I define the monthly log-spread innovation as
\begin{equation}
x_{c,t}
=
\Delta \log s_{c,t}
=
\log s_{c,t}-\log s_{c,t-1}.
\label{eq:method_x}
\end{equation}
For each country, I compute the full-sample standard deviation
$\hat{\sigma}_c=\operatorname{sd}(x_{c,t})$. A country-level stress event is
identified when the positive log-spread innovation exceeds a country-specific
volatility threshold:
\begin{equation}
I_{c,t}(\sigma)
=
\mathbf{1}\left[x_{c,t}>\sigma\hat{\sigma}_c\right].
\label{eq:method_hit}
\end{equation}
The baseline specification uses $\sigma=1.5$. Robustness exercises repeat the
analysis for
\[
\sigma\in\{1.25,1.50,1.75,2.00,2.25,2.50\}.
\]
This country-normalized threshold avoids imposing a common basis-point
definition of stress on sovereigns with very different spread levels and
volatility regimes.

The sovereign-stress avalanche size in month $t$ is defined as the number of
countries simultaneously in stress:
\begin{equation}
S_t(\sigma)
=
\sum_{c=1}^{n} I_{c,t}(\sigma).
\label{eq:method_avalanche}
\end{equation}
Thus, $S_t=1$ corresponds to an isolated country-level stress event, while
large values of $S_t$ indicate synchronized regional stress. In the empirical
application, $n=11$, so avalanche sizes are mechanically bounded between zero
and eleven. This finite support is important for interpretation: the analysis
tests for finite-size criticality signatures rather than for an asymptotic
infinite-system power law.

\subsection{Power-law estimation and tail diagnostics}
\label{subsec:tail_diagnostics}

The paper studies three related statistical objects: avalanche sizes
$S_t$, absolute monthly spread changes $|\Delta s_{c,t}|$, and inter-event
times $\tau_{c,k}$ between consecutive stress events for country $c$. For each
object, I estimate the power-law exponent $\hat{\alpha}$ and lower cutoff
$x_{\min}$ using the maximum-likelihood procedure of \citet{Clauset2009}. The
cutoff $x_{\min}$ is chosen by minimizing the Kolmogorov--Smirnov distance
between the empirical and fitted complementary cumulative distribution
functions.

The fitted power law is then compared with lognormal alternatives using the
non-nested likelihood-ratio framework of \citet{Vuong1989}. I report the
normalized log-likelihood ratio $R_{\mathrm{PL/LN}}$ and its associated
$p$-value. Positive values of $R_{\mathrm{PL/LN}}$ favor the power law, while
negative values favor the lognormal. Following \citet{StumpfPorter2012}, I do
not interpret the analysis as a mechanical test of a pure asymptotic power
law. Instead, the relevant question is whether the empirical distributions
belong to a heavy-tail boundary regime in which Gaussian or exponential
benchmark models are inappropriate and the power-law and lognormal
specifications are difficult to separate decisively in finite samples.

The main object for the SOC interpretation is the avalanche-size distribution,
not the marginal distribution of individual spread changes. Individual
monthly spread changes may be influenced by liquidity shocks, rating news,
country-specific jumps, and global risk appetite. Avalanche sizes instead
measure the cross-sectional synchronization of stress events, which is the
relevant object for detecting regional propagation.

\subsection{Filtered sovereign-spread networks}
\label{subsec:filtered_networks}

To study network amplification, I construct rolling networks of sovereign
spread co-movement using 24-month windows of $x_{c,t}=\Delta\log s_{c,t}$.
For each window ending in month $t$, three network filters are computed over
the common set of eleven countries.

The first network is the absolute-correlation network, $G_t^{\rho}$. An
undirected edge $(i,j)$ is included when
\begin{equation}
|\hat{\rho}_{ij,t}|>\rho^{*},
\qquad
\rho^{*}=0.4,
\label{eq:corr_threshold}
\end{equation}
where $\hat{\rho}_{ij,t}$ is the sample correlation between spread innovations
of sovereigns $i$ and $j$ within the rolling window. Edge weights are given by
$w_{ij,t}=|\hat{\rho}_{ij,t}|$.

The second network is the partial-correlation network, $G_t^{\mathrm{part}}$.
Let $\Sigma_t$ denote the covariance matrix of spread innovations in the
rolling window and let $\Omega_t=\Sigma_t^{-1}$ denote the corresponding
precision matrix. The partial correlation between sovereigns $i$ and $j$ is
defined as
\begin{equation}
\hat{\rho}^{\mathrm{part}}_{ij,t}
=
-
\frac{\Omega_{ij,t}}
{\sqrt{\Omega_{ii,t}\Omega_{jj,t}}}.
\label{eq:method_partial_corr}
\end{equation}
An edge is included when
$|\hat{\rho}^{\mathrm{part}}_{ij,t}|>\rho^{*}$, with weight
$|\hat{\rho}^{\mathrm{part}}_{ij,t}|$. This network is intended to reduce the
influence of common linear dependence and to isolate conditional co-movement
among sovereigns.

The third network is the minimum spanning tree, $G_t^{\mathrm{MST}}$, computed
from correlation distances
\begin{equation}
d_{ij,t}=1-|\hat{\rho}_{ij,t}|.
\label{eq:mst_distance}
\end{equation}
Following the filtering logic of \citet{Mantegna1999} and
\citet{Tumminello2005}, the MST preserves the strongest hierarchical
co-movement structure while imposing a fixed number of edges. Since the system
contains $n=11$ sovereigns, each MST contains exactly $n-1=10$ edges.

The correlation network captures total co-movement, the partial-correlation
network captures conditional co-movement, and the MST provides a sparse
topological benchmark. The comparison across these three filters is important
because raw correlations may become mechanically dense during high-volatility
episodes, whereas filtered networks provide a more conservative view of
regional stress transmission.

\subsection{Network amplification and geometry}
\label{subsec:network_geometry}

For each network snapshot, I compute density, the Herfindahl--Hirschman index
of node strengths, spectral amplification measures, and curvature-based
geometric summaries. Let $W_t=[w_{ij,t}]$ denote the weighted adjacency matrix
of a given network at time $t$. Network density is the realized share of
possible undirected edges, while the HHI of node strengths is defined as
\begin{equation}
HHI_t
=
\sum_{c=1}^{n}
\left(
\frac{\sum_j w_{cj,t}}
{\sum_i\sum_j w_{ij,t}}
\right)^2.
\label{eq:method_hhi}
\end{equation}
A lower HHI indicates that network strength is more evenly distributed across
sovereigns.

The main amplification measures are normalized spectral radii. I use two
normalizations:
\begin{equation}
\rho^{\infty}_t
=
\lambda_{\max}
\left(
\frac{W_t}{\|W_t\|_{\infty}}
\right),
\qquad
\rho^{F}_t
=
\lambda_{\max}
\left(
\frac{W_t}{\|W_t\|_{F}}
\right),
\label{eq:method_spectral}
\end{equation}
where $\|W_t\|_{\infty}$ is the maximum row-sum norm and $\|W_t\|_F$ is the
Frobenius norm. These normalizations avoid the mechanical unit eigenvalue that
would arise from row-stochastic normalization. The Spectral Fragility Index is
defined as
\begin{equation}
SFI_t
=
\frac{\rho_t}{1-\rho_t},
\label{eq:method_sfi}
\end{equation}
where $\rho_t$ denotes either normalized spectral radius. The index increases
nonlinearly as the network approaches the high-amplification boundary
$\rho_t\rightarrow 1$.

I also compute Forman--Ricci and Ollivier--Ricci curvature measures. Mean
Forman--Ricci curvature, $\bar{\kappa}^{F}_t$, is computed using Equation
\eqref{eq:forman}. Mean Ollivier--Ricci curvature, $\bar{\kappa}^{O}_t$, is
computed through an optimal-transport problem between neighborhood measures,
as in Equation \eqref{eq:ollivier}. In the empirical interpretation, curvature
is treated as an auxiliary measure of geometric integration. The primary
fragility indicators are the spectral radius, SFI, density, and HHI, because
these quantities map more directly into amplification and synchronization.

\subsection{Placebo synchronization test}
\label{subsec:placebo}

A key concern is that large avalanches may arise mechanically from
country-level stress frequencies. To address this concern, I implement a
country-level reshuffling placebo. For each replication $b=1,\ldots,B$, with
$B=1{,}000$, the stress indicator sequence
$I_{c,t}(\sigma)$ is independently shuffled across calendar months within
each country. This procedure preserves the number of stress events for each
sovereign, but destroys cross-country timing.

For each placebo replication, I recompute the avalanche-size process
$S_t^{(b)}$ and four synchronization statistics:
\[
\max_t S_t^{(b)}, \qquad
\mathbb{E}[S_t^{(b)}\mid S_t^{(b)}>0], \qquad
\Pr(S_t^{(b)}\geq 4), \qquad
\Pr(S_t^{(b)}\geq 6).
\]
The placebo $p$-value is the share of replications in which the placebo
statistic equals or exceeds its observed counterpart. This test directly asks
whether the observed regional synchronization is stronger than what would be
expected from independent country-level stress timing.

\subsection{Lead-lag network association}
\label{subsec:lead_lag}

Finally, I evaluate whether network amplification contains early-warning
information about subsequent regional stress. For each network metric $m_t$
computed at snapshot $t$, I define the future three-month avalanche outcome as
\begin{equation}
Y_t
=
\max_{u\in(t,t+3]} S_u.
\label{eq:future_avalanche}
\end{equation}
I then compute Pearson correlations and simple heteroskedasticity-robust OLS
associations between $m_t$ and $Y_t$. These regressions are not interpreted as
causal estimates. Their purpose is to assess whether contemporaneous network
states are informative about near-future regional stress intensity. The main
metrics of interest are the spectral radius, SFI, network density, and HHI,
especially in the partial-correlation network.

\section{Empirical Results}
\label{sec:results}

This section reports the empirical evidence in six steps, following the order
of hypotheses stated in Section~\ref{sec:theory}. I first evaluate whether
sovereign-stress avalanche sizes exhibit heavy-tailed behavior (H1). I then
test whether the observed synchronization is stronger than what would be
expected from country-level stress frequencies alone (H2). I next examine
threshold robustness (H3), country participation heterogeneity (H4), network
amplification during large avalanche regimes (H5), and the lead--lag
association between network states and subsequent regional stress (H6).

\subsection{Avalanche-size heavy tails (H1)}
\label{subsec:results_heavytails}

Table~\ref{tab:tail_diagnostics} reports formal tail diagnostics for three
statistical objects: avalanche sizes $S_t$, absolute monthly EMBI spread
changes $|\Delta s_{c,t}|$, and inter-event times $\tau_{c,k}$ between
consecutive stress events of country $c$. The baseline avalanche-size
distribution yields an estimated exponent of $\hat{\alpha}=1.77$ with
$x_{\min}=1$. The likelihood-ratio comparison against the lognormal
alternative is negative, $R_{\mathrm{PL/LN}}=-1.61$, but not statistically
decisive at conventional levels ($p_{LN}=0.107$). The evidence does not
uniquely identify a pure power law, but it is consistent with a finite-size
heavy-tailed avalanche distribution. The number of avalanche events above the
estimated cutoff ($n=40$) is below the threshold of $n\geq 50$ that
\citet{Clauset2009} recommend for stable maximum-likelihood
estimation, which provides an additional reason to interpret the exponent
estimate cautiously.

Pooled absolute monthly spread changes also display heavy-tailed behavior,
with $\hat{\alpha}=2.37$ and $x_{\min}\simeq 86$ basis points. The comparison
against the lognormal is again not decisive
($R_{\mathrm{PL/LN}}=-0.76$, $p_{LN}=0.448$), although the exponential
alternative is strongly rejected ($R_{\mathrm{PL/Exp}}=3.43$, $p=0.001$).
Inter-event times produce a similar conclusion, with $\hat{\alpha}=2.38$,
$x_{\min}=12$ months, and no decisive likelihood-ratio separation from the
lognormal. Taken together, the results place the EMBI system in the
heavy-tail boundary regime emphasized by \citet{StumpfPorter2012}: the
relevant empirical claim is not that a pure asymptotic power law is proven,
but that avalanche sizes, spread changes, and waiting times are all far from
thin-tailed benchmark behavior.

\begin{table}[!htbp]
\centering
\footnotesize
\captionsetup{width=0.95\textwidth,justification=centering}

\begin{threeparttable}
\caption{Tail diagnostics for SOC-relevant statistical objects in LATAM EMBI data}
\label{tab:tail_diagnostics}

\centering

\textbf{Panel A. Power-law tail estimates}

\vspace{0.4em}

\begin{tabular*}{0.80\textwidth}{@{\extracolsep{\fill}}lrrrr}
\toprule
Object & $n$ & $x_{\min}$ & $\hat{\alpha}$ & KS $D$ \\
\midrule
Avalanche size $S_t$ 
& 40 & 1.000 & 1.771 & 0.095 \\

$|\Delta s_{c,t}|$ pooled 
& 2442 & 85.459 & 2.372 & 0.028 \\

Inter-event times (months) 
& 129 & 11.991 & 2.381 & 0.084 \\
\bottomrule
\end{tabular*}

\vspace{1.1em}

\textbf{Panel B. Likelihood-ratio comparisons and interpretation}

\vspace{0.4em}

\begin{tabular*}{0.95\textwidth}{@{\extracolsep{\fill}}lrrrrp{0.23\textwidth}}
\toprule
Object 
& $R_{\mathrm{PL/LN}}$ 
& $p_{\mathrm{LN}}$ 
& $R_{\mathrm{PL/Exp}}$ 
& $p_{\mathrm{Exp}}$ 
& Verdict \\
\midrule

Avalanche size $S_t$ 
& -1.612 & 0.107 & -0.502 & 0.616 
& Heavy-tail boundary; PL not decisively rejected vs.\ LN \\

$|\Delta s_{c,t}|$ pooled 
& -0.758 & 0.448 & 3.426 & 0.001 
& Heavy-tail boundary; PL favored over Exp \\

Inter-event times (months) 
& -0.606 & 0.544 & 1.106 & 0.269 
& Heavy-tail boundary; PL not decisively rejected vs.\ LN \\

\bottomrule
\end{tabular*}

\vspace{0.4em}

\begin{tablenotes}\footnotesize
\item $\hat{\alpha}$ and $x_{\min}$ are maximum-likelihood estimates of the power-law tail exponent and lower cutoff, respectively, following \citet{Clauset2009}. KS $D$ is the Kolmogorov--Smirnov distance from the fitted power law above $x_{\min}$.
\item $R_{\mathrm{PL/LN}}$ and $R_{\mathrm{PL/Exp}}$ are Vuong normalized log-likelihood ratios comparing the power-law model against lognormal and exponential alternatives, respectively. Positive values favor the power-law model.
\item The results support a conservative heavy-tail boundary interpretation in the sense of \citet{StumpfPorter2012}: power-law and lognormal alternatives are difficult to discriminate in finite samples, while the exponential benchmark is clearly rejected only for pooled absolute spread changes.
\item The avalanche-size evidence should therefore be interpreted as finite-size criticality evidence, not as proof of a unique asymptotic power law.
\end{tablenotes}

\end{threeparttable}
\end{table}

\begin{figure}[H]
\centering
\includegraphics[width=\linewidth]{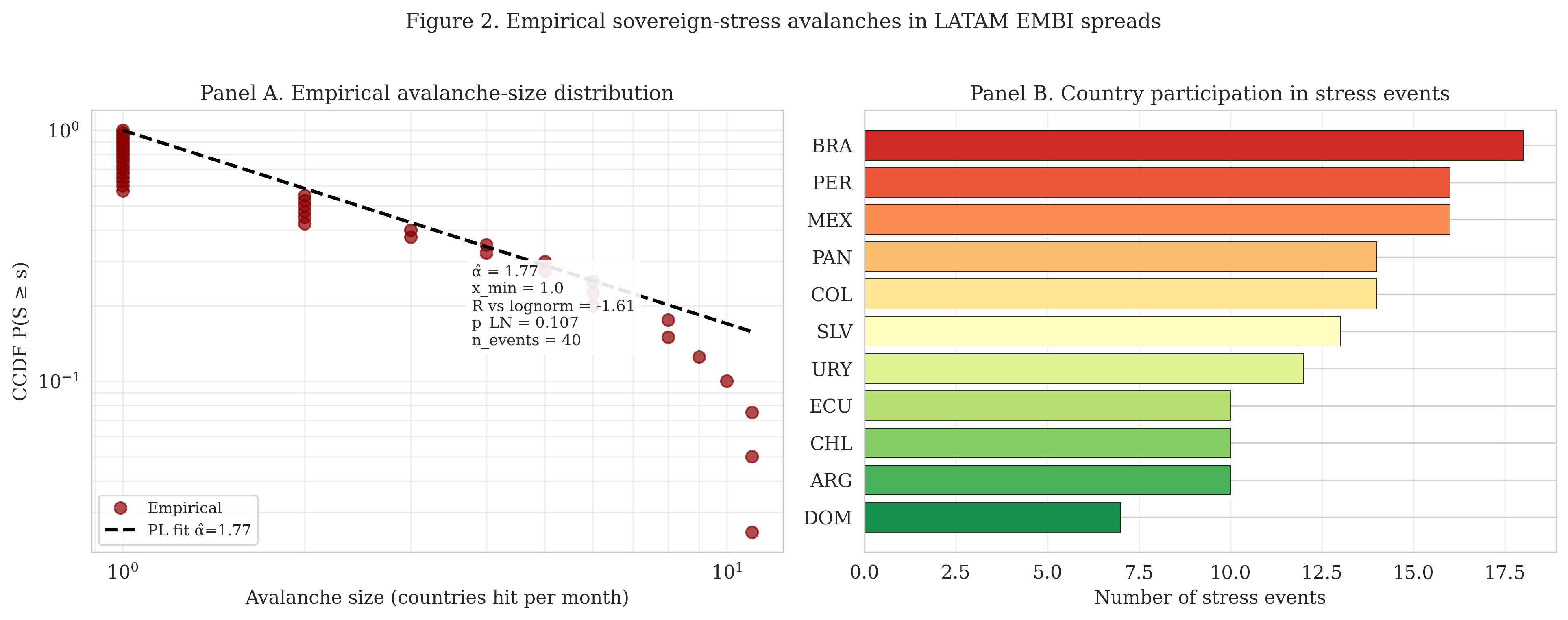}
\caption{Empirical sovereign-stress avalanches in LATAM EMBI spreads.
Panel~A reports the avalanche-size CCDF with maximum-likelihood power-law
fit ($\hat{\alpha}=1.77$). Panel~B reports country participation by
stress-event count.}
\label{fig:avalanche_soc}
\end{figure}

Figure~\ref{fig:avalanche_soc} displays the empirical avalanche-size CCDF and
the country participation ranking. Panel~A shows that most stress events are
small, but the upper tail includes regional avalanches involving a large
share of the system. Panel~B shows that participation is highly heterogeneous
across sovereigns, a result analyzed further in
Section~\ref{subsec:results_participation}.

\begin{figure}[H]
\centering
\includegraphics[width=\linewidth]{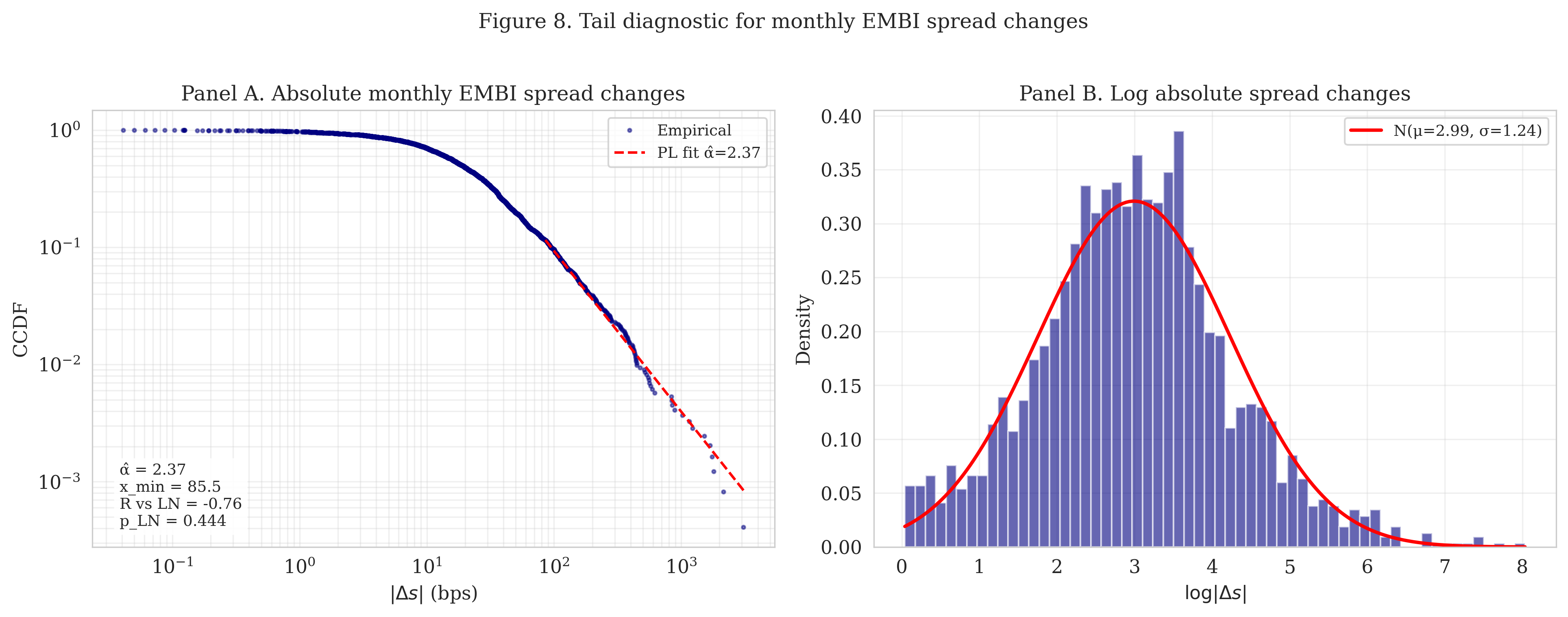}
\caption{Tail diagnostic for monthly EMBI spread changes. Panel~A reports
pooled absolute monthly spread changes with power-law fit
($\hat{\alpha}=2.37$, $x_{\min}\simeq 86$ basis points). Panel~B reports
the histogram of log-absolute spread changes with a normal overlay.}
\label{fig:dspread_diagnostic}
\end{figure}

Figure~\ref{fig:dspread_diagnostic} reports the corresponding diagnostic for
pooled absolute monthly spread changes. This figure is not the primary SOC
object, because the relevant propagation object is the synchronized avalanche
size $S_t$. It is included to show that the underlying EMBI innovations are
also heavy-tailed.

The estimated avalanche exponent lies in the range commonly associated with
finite-size critical systems \citep{Bak1996, NewmanPL2005}. However, because
the lognormal alternative cannot be decisively ruled out and the number of
tail observations is moderate, the interpretation is deliberately
conservative: the evidence supports heavy-tailed finite-size avalanche
behavior, not a claim of uniquely identified asymptotic self-organized
criticality. \textbf{H1 is therefore confirmed in its conservative form: the
distribution of sovereign-stress avalanche sizes is heavy-tailed and lies in
a finite-size criticality regime, with an estimated exponent in the plausible
critical range and lognormal alternatives not decisively preferred.}

\subsection{Non-random synchronization (H2)}
\label{subsec:results_placebo}

Table~\ref{tab:placebo} reports the country-level reshuffling placebo test.
The placebo preserves each sovereign's total number of stress events but
randomizes the timing of those events within country. It therefore tests
whether the observed avalanche sizes are mechanically generated by
country-specific hit frequencies.

The observed maximum avalanche size is $\max_t S_t=11$, whereas the placebo
distribution has a mean maximum size of approximately $3.47$ and a $95$th
percentile equal to $4$. The observed share of large avalanches is also far
above its placebo counterpart. All four synchronization statistics reject the
reshuffled-timing null at $p<0.001$. \textbf{This is the strongest result of
the paper: regional EMBI stress synchronization is not an artifact of having
several volatile countries in the same sample, and H2 is confirmed
decisively.}

\begin{table}[!htbp]
\centering
\footnotesize
\begin{threeparttable}
\caption{Placebo test of cross-country stress synchronization. Country-level stress indicators are independently reshuffled across calendar months, destroying cross-country timing dependence while preserving country-level event frequencies.}
\label{tab:placebo}
\begin{tabular}{lrrrr}
\toprule
Statistic & Observed & Placebo mean & Placebo P95 & $p$-value \\
\midrule
Max avalanche size $S_{\max}$ & 11 & 3.47 & 4.00 & $<0.001$ \\
Mean positive avalanche size $\mathbb{E}[S\mid S>0]$ & 3.50 & 1.32 & 1.40 & $<0.001$ \\
Share $P(S\geq 4)$ & 0.350 & 0.005 & 0.019 & $<0.001$ \\
Share $P(S\geq 6)$ & 0.250 & 0.000 & 0.000 & $<0.001$ \\
\bottomrule
\end{tabular}
\begin{tablenotes}\footnotesize
\item Placebo replications: $B=1{,}000$, seed $=2026$.
\item The reshuffling placebo preserves each sovereign's total number of stress events but destroys the common calendar timing of those events.
\item Under independent country-level stress timing, the mean placebo maximum avalanche size is $3.47$ and the 95th percentile is $4$. The observed value $S_{\max}=11$ is not reached in any of the 1{,}000 placebo replications.
\item All four synchronization statistics reject the independent-timing null at $p<0.001$, confirming that simultaneous LATAM sovereign stress is a non-trivial multi-country phenomenon.
\end{tablenotes}
\end{threeparttable}
\end{table}

It is important to be precise about what the placebo identifies. By
construction, the test rules out the null in which country-level stress
events are timed independently across sovereigns. It therefore demonstrates
that the observed cross-sectional clustering is non-random conditional on
country-specific event frequencies. The test does not separate two
substantive sources of synchronization: endogenous regional propagation
(stress in country $i$ raising the probability of stress in country $j$) and
exogenous common-factor synchronization (stress in country $i$ and country
$j$ jointly responding to a global shock). The largest avalanches in the
sample coincide with global systemic events, which suggests that
common-factor synchronization plays a non-trivial role. The interpretation
of the paper is consistent with this observation: the avalanche framework
documents the empirical existence of synchronized regional stress as a
finite-size statistical phenomenon, without claiming to identify a unique
structural transmission channel.

Table~\ref{tab:top_avalanches} lists the largest avalanche months and their
historical interpretation. The three avalanches with $S_t=11$ coincide with
global systemic events: the Lehman collapse, the post-Lehman emerging-market
rout, and the COVID-19 global financial shock. More broadly, the largest
avalanches align with recognizable international stress episodes rather than
with isolated country-specific spread events. This historical coherence
supports the interpretation of $S_t$ as a meaningful regional stress
variable.

\begin{table}[!htbp]
\centering
\footnotesize
\begin{threeparttable}
\caption{Top 12 largest empirical avalanches in LATAM EMBI spreads, with historical context.}
\label{tab:top_avalanches}
\begin{tabular}{rlrp{0.30\linewidth}p{0.32\linewidth}}
\toprule
Rank & Month & $S_t$ & Countries hit & Historical context \\
\midrule
1 & 2020-03 & 11 & ARG, BRA, CHL, COL, DOM, ECU, MEX, PAN, PER, SLV, URY & COVID-19 global rout, Ecuador default \\
2 & 2008-09 & 11 & ARG, BRA, CHL, COL, DOM, ECU, MEX, PAN, PER, SLV, URY & Lehman collapse (global GFC peak) \\
3 & 2008-10 & 11 & ARG, BRA, CHL, COL, DOM, ECU, MEX, PAN, PER, SLV, URY & Post-Lehman EM rout \\
4 & 2007-11 & 10 & ARG, BRA, CHL, COL, DOM, MEX, PAN, PER, SLV, URY & Pre-crisis volatility wave (Bear Stearns hedge funds collapse) \\
5 & 2011-09 & 9 & ARG, BRA, CHL, COL, MEX, PAN, PER, SLV, URY & Eurozone crisis intensifies \\
6 & 2010-05 & 8 & BRA, CHL, COL, DOM, MEX, PER, SLV, URY & Eurozone crisis (first Greek bailout) \\
7 & 2012-05 & 8 & ARG, BRA, CHL, COL, MEX, PAN, PER, URY & Greek collapse round II \\
8 & 2008-06 & 6 & BRA, COL, MEX, PAN, PER, SLV & Subprime spillover to EM \\
9 & 2020-02 & 6 & CHL, ECU, MEX, PAN, PER, URY & COVID-19 onset, Ecuador default brewing \\
10 & 2014-01 & 6 & ARG, BRA, COL, MEX, PER, URY & EM Fragile Five sell-off \\
11 & 2011-08 & 5 & BRA, CHL, COL, MEX, PAN & U.S. debt-ceiling / S\&P downgrade \\
12 & 2022-06 & 5 & BRA, COL, DOM, ECU, MEX & Fed front-loading; LATAM commodity stress \\
\bottomrule
\end{tabular}
\begin{tablenotes}\footnotesize
\item Each row is a calendar month ranked by $S_t$ = number of countries simultaneously exceeding their $1.5\sigma$ threshold.
\item Every avalanche of size $S\geq 9$ coincides with a documented global systemic event.
\item All $S=11$ months are universal-stress events (every LATAM sovereign in the panel hit simultaneously).
\end{tablenotes}
\end{threeparttable}
\end{table}

\subsection{Threshold robustness (H3)}
\label{subsec:results_threshold}

\begin{figure}[H]
\centering
\includegraphics[width=0.9\linewidth]{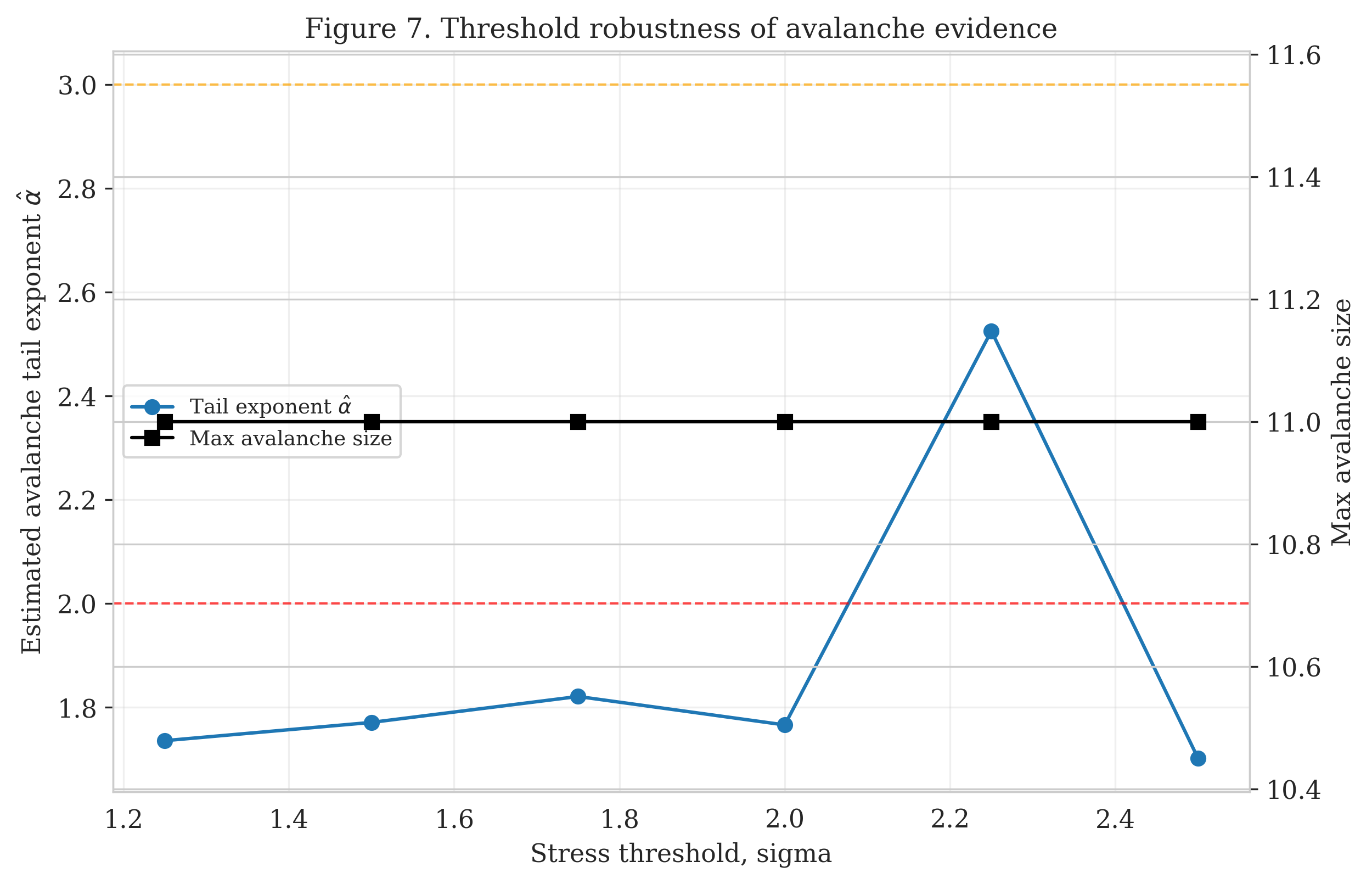}
\caption{Threshold robustness of avalanche evidence. The blue line reports
the estimated avalanche tail exponent $\hat{\alpha}$ on the left axis. The
black line reports the maximum avalanche size on the right axis. Thresholds
range from $\sigma=1.25$ to $\sigma=2.50$.}
\label{fig:robustness}
\end{figure}

Figure~\ref{fig:robustness} and Table~\ref{tab:threshold_robustness} examine
whether the avalanche evidence depends on the baseline stress threshold
$\sigma=1.5$. The estimated avalanche exponent remains in a narrow range for
the empirically informative thresholds:
\[
\hat{\alpha}\in[1.70,1.83]
\quad \text{for} \quad
\sigma\leq 2.00.
\]
At higher thresholds, the number of avalanche months falls mechanically and
tail estimates become less stable, as expected in a finite system.
Importantly, the maximum avalanche size remains equal to eleven at every
threshold in the grid. The largest regional stress events are therefore not
artifacts of the baseline threshold choice.

\begin{table}[!htbp]
\centering
\footnotesize
\begin{threeparttable}
\caption{Sensitivity of avalanche evidence to the stress-threshold parameter $\sigma$.}
\label{tab:threshold_robustness}
\begin{tabular*}{\textwidth}{@{\extracolsep{\fill}}rrrrrrrrrr}
\toprule
$\sigma$ & Country hits & Avalanche months & Mean $S$ & Max $S$ & $P(S\geq 6)$ & $\hat{\alpha}$ & $x_{\min}$ & $R_{\mathrm{PL/LN}}$ & $p_{\mathrm{LN}}$ \\
\midrule
1.25 & 199 & 54 & 3.69 & 11 & 0.241 & 1.736 & 1.000 & -2.106 & 0.035 \\
1.50 & 140 & 40 & 3.50 & 11 & 0.250 & 1.771 & 1.000 & -1.612 & 0.107 \\
1.75 & 112 & 35 & 3.20 & 11 & 0.171 & 1.821 & 1.000 & -1.393 & 0.164 \\
2.00 & 67 & 19 & 3.53 & 11 & 0.263 & 1.766 & 1.000 & -1.181 & 0.238 \\
2.25 & 52 & 13 & 4.00 & 11 & 0.308 & 2.525 & 4.000 & -1.005 & 0.315 \\
2.50 & 41 & 10 & 4.10 & 11 & 0.300 & 1.702 & 1.000 & -0.878 & 0.380 \\
\bottomrule
\end{tabular*}
\begin{tablenotes}\footnotesize
\item Avalanche months are months with $S_t>0$. Mean $S$ denotes the average avalanche size conditional on an avalanche occurring.
\item The main qualitative conclusion is robust across thresholds: universal avalanches with $S_{\max}=11$ appear at every tested value of $\sigma$.
\item Tail-exponent stability is strongest over the empirically informative range $\sigma\leq2.00$, where $\hat{\alpha}\in[1.70,1.83]$.
\item At higher thresholds, the number of avalanche months falls mechanically and tail estimates become less stable, as expected in a finite-size system.
\item The likelihood-ratio evidence should be interpreted conservatively: at $\sigma=1.25$, the lognormal alternative is favored at conventional significance levels, while for the baseline and higher thresholds it is not decisively preferred.
\end{tablenotes}
\end{threeparttable}
\end{table}

The robustness exercise supports the finite-size interpretation. The precise
tail exponent varies with the threshold, but the existence of very large
regional avalanches does not. This distinction matters because the central
empirical claim is not tied to one arbitrary cutoff; it is the persistence of
large synchronized events across reasonable event definitions. \textbf{H3 is
therefore confirmed: the estimated tail exponent and maximum avalanche size
are stable across the empirically informative threshold range.}

\subsection{Heterogeneous country participation (H4)}
\label{subsec:results_participation}

Table~\ref{tab:country_thresholds} reports the country-specific stress
thresholds and event counts. Stress participation is highly heterogeneous.
Brazil records the largest number of stress events, followed by Mexico and
Peru, while the Dominican Republic records the fewest. A $\chi^2$ test
against uniform participation rejects equality of stress frequencies across
countries, indicating that avalanche participation is not evenly distributed
across the regional system.

\begin{table}[!htbp]
\centering
\scriptsize
\setlength{\tabcolsep}{3pt}
\renewcommand{\arraystretch}{1.08}
\begin{threeparttable}
\caption{Country-specific stress thresholds and observed event counts, $\sigma=1.5$ specification}
\label{tab:country_thresholds}

\begin{tabular}{llrrrrC{1.15cm}C{1.15cm}C{1.15cm}}
\toprule
Code 
& Country 
& $\sigma_{\Delta\log s}$ 
& Threshold 
& $N_c$ 
& Hit rate 
& Mean $|\Delta s|$ 
& P95 $|\Delta s|$ 
& Max $|\Delta s|$ \\
&&&&&& (bp) & (bp) & (bp) \\
\midrule
BRA & Brazil      & 0.111 & 0.166 & 18 & 0.081 & 22  & 63  & 138 \\
MEX & Mexico      & 0.109 & 0.163 & 16 & 0.072 & 22  & 62  & 281 \\
PER & Peru        & 0.135 & 0.203 & 16 & 0.072 & 20  & 50  & 203 \\
COL & Colombia    & 0.131 & 0.197 & 14 & 0.063 & 24  & 61  & 213 \\
PAN & Panama      & 0.129 & 0.193 & 14 & 0.063 & 20  & 54  & 207 \\
SLV & El Salvador & 0.119 & 0.179 & 13 & 0.059 & 53  & 178 & 431 \\
URY & Uruguay     & 0.136 & 0.203 & 12 & 0.054 & 22  & 63  & 421 \\
ARG & Argentina   & 0.166 & 0.249 & 10 & 0.045 & 132 & 441 & 1,751 \\
CHL & Chile       & 0.117 & 0.175 & 10 & 0.045 & 15  & 37  & 142 \\
ECU & Ecuador     & 0.195 & 0.293 & 10 & 0.045 & 153 & 448 & 3,087 \\
DOM & Dom. Rep.   & 0.119 & 0.178 & 7  & 0.032 & 37  & 91  & 1,018 \\
\bottomrule
\end{tabular}

\begin{tablenotes}\footnotesize
\item Notes: A stress event for country $c$ at month $t$ is defined by
$\Delta\log s_{c,t} > \sigma\cdot\hat{\sigma}_c$, where $\hat{\sigma}_c$ is
the country-specific standard deviation of monthly log changes.
\item Countries with the largest tail jumps, especially Argentina and Ecuador,
display fewer events because their volatility is concentrated in a small number
of regime-defining episodes.
\item Mid-credibility and highly traded sovereigns, especially Brazil, Mexico,
Peru, Colombia, and Panama, accumulate the largest event counts.
\end{tablenotes}

\end{threeparttable}
\end{table}

The substantive result is that avalanche participation is not mechanically
ordered by average spread level. Argentina and Ecuador are among the
highest-spread sovereigns in the panel, but they do not dominate the
stress-event count. By contrast, more liquid and financially integrated
sovereigns such as Brazil, Mexico, Peru, Colombia, and Panama participate
more frequently in regional stress episodes. This pattern is consistent with
a network-amplification interpretation: the most frequently activated nodes
are not necessarily the weakest sovereigns, but those that are sufficiently
liquid, investable, and integrated into the regional emerging-market debt
portfolio to react repeatedly to changes in global risk conditions.

\begin{figure}[H]
\centering
\includegraphics[width=\linewidth]{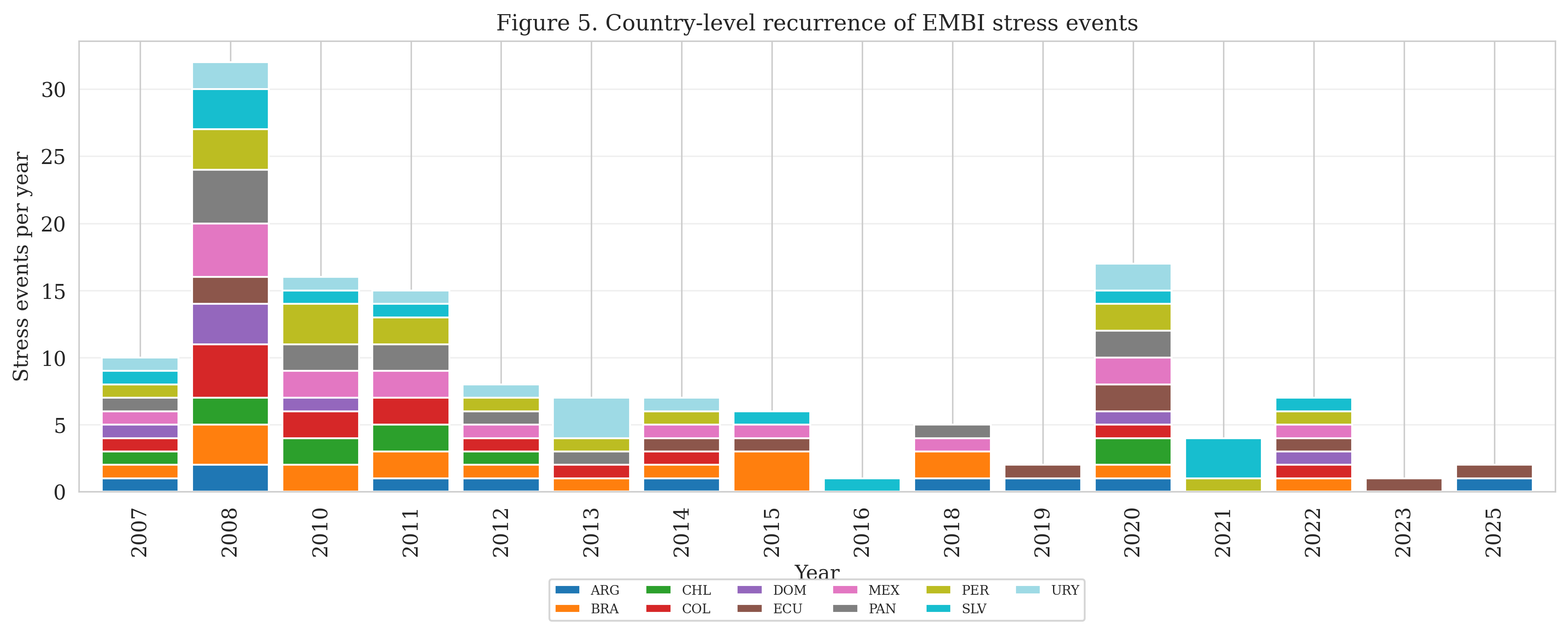}
\caption{Country-level recurrence of EMBI stress events by year, stacked
across countries. The largest cluster occurs in 2008, followed by stress
clusters around 2010--2011 and 2020.}
\label{fig:recurrence}
\end{figure}

\begin{figure}[H]
\centering
\includegraphics[width=0.75\linewidth]{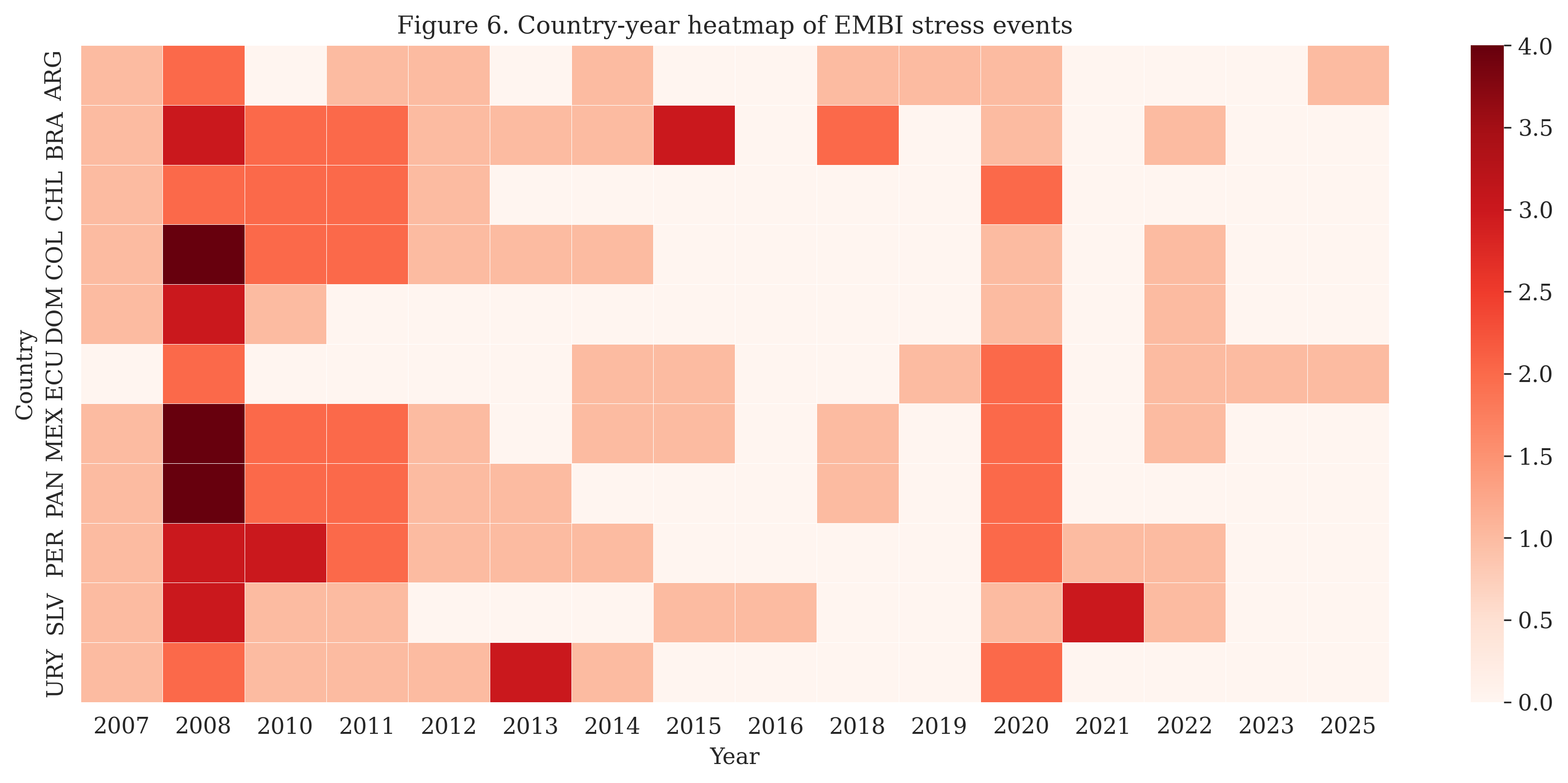}
\caption{Country-year heatmap of EMBI stress events. Cell intensity reports
the number of months in which country $c$ exceeded its $1.5\sigma$ stress
threshold in year $y$.}
\label{fig:heatmap}
\end{figure}

Figures~\ref{fig:recurrence} and~\ref{fig:heatmap} display the timing of
country-level stress recurrence. The largest clusters occur around the global
financial crisis and the COVID-19 shock, with additional stress episodes in
2010--2011 and during selected emerging-market repricing periods. The
temporal pattern reinforces the interpretation that country participation is
shaped by regional and global synchronization, not only by idiosyncratic
fiscal weakness. \textbf{H4 is therefore confirmed: country participation is
heterogeneous and is not mechanically ordered by average spread level.}

\subsection{Network amplification during large avalanches (H5)}
\label{subsec:results_network_regime}

\begin{figure}[H]
\centering
\includegraphics[width=0.85\linewidth]{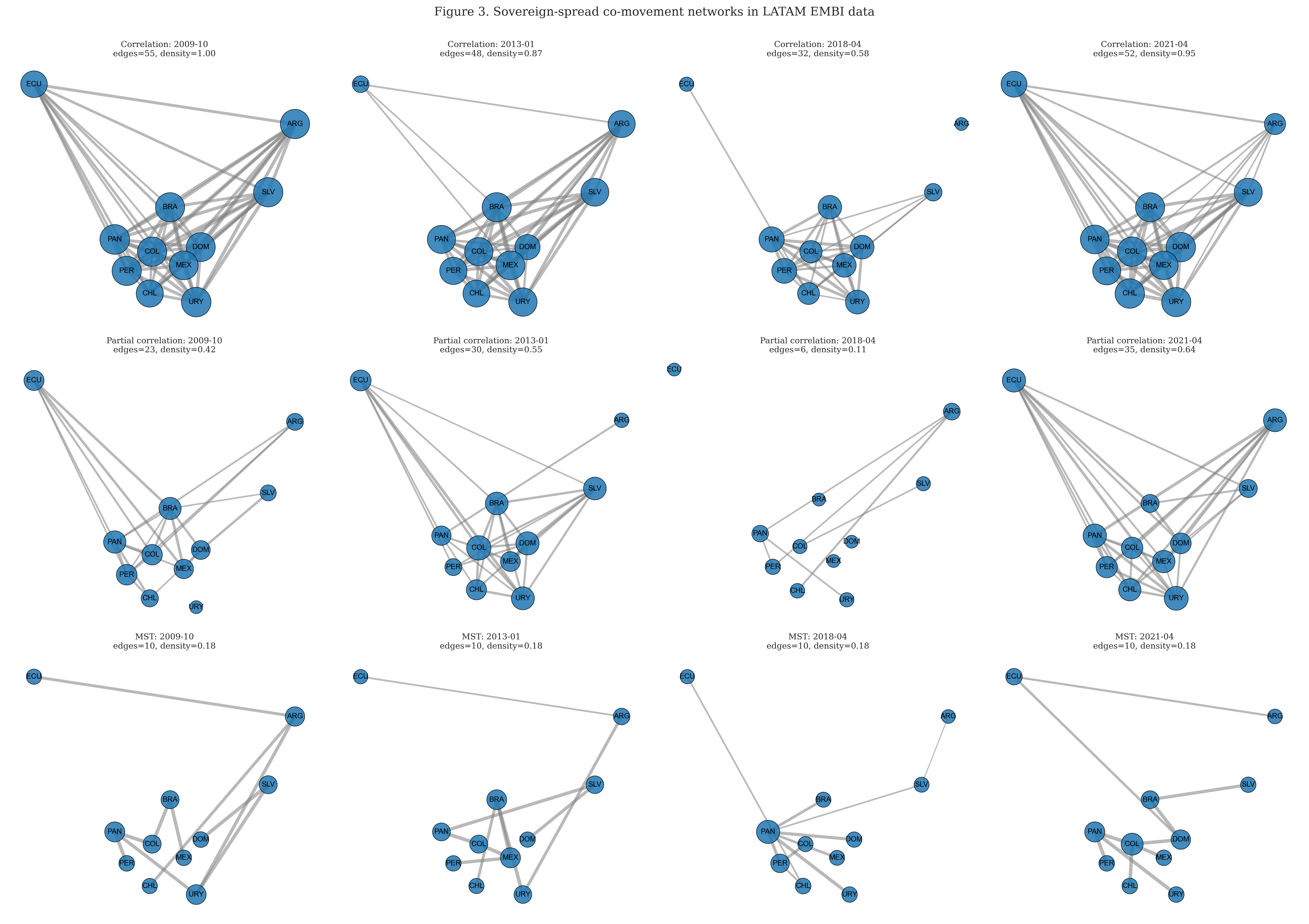}
\caption{Sovereign-spread co-movement networks in LATAM EMBI data at four
representative dates. Top row: correlation network. Middle row:
partial-correlation network. Bottom row: minimum spanning tree.}
\label{fig:networks}
\end{figure}

Figure~\ref{fig:networks} displays the three filtered co-movement networks at
four representative dates. The correlation network becomes extremely dense in
systemic periods, reflecting common regional repricing. The
partial-correlation network is sparser and more informative about conditional
co-movement after filtering common linear dependence. The minimum spanning
tree provides a fixed-edge benchmark for the strongest hierarchical
co-movement structure.

\begin{figure}[H]
\centering
\includegraphics[width=0.75\linewidth]{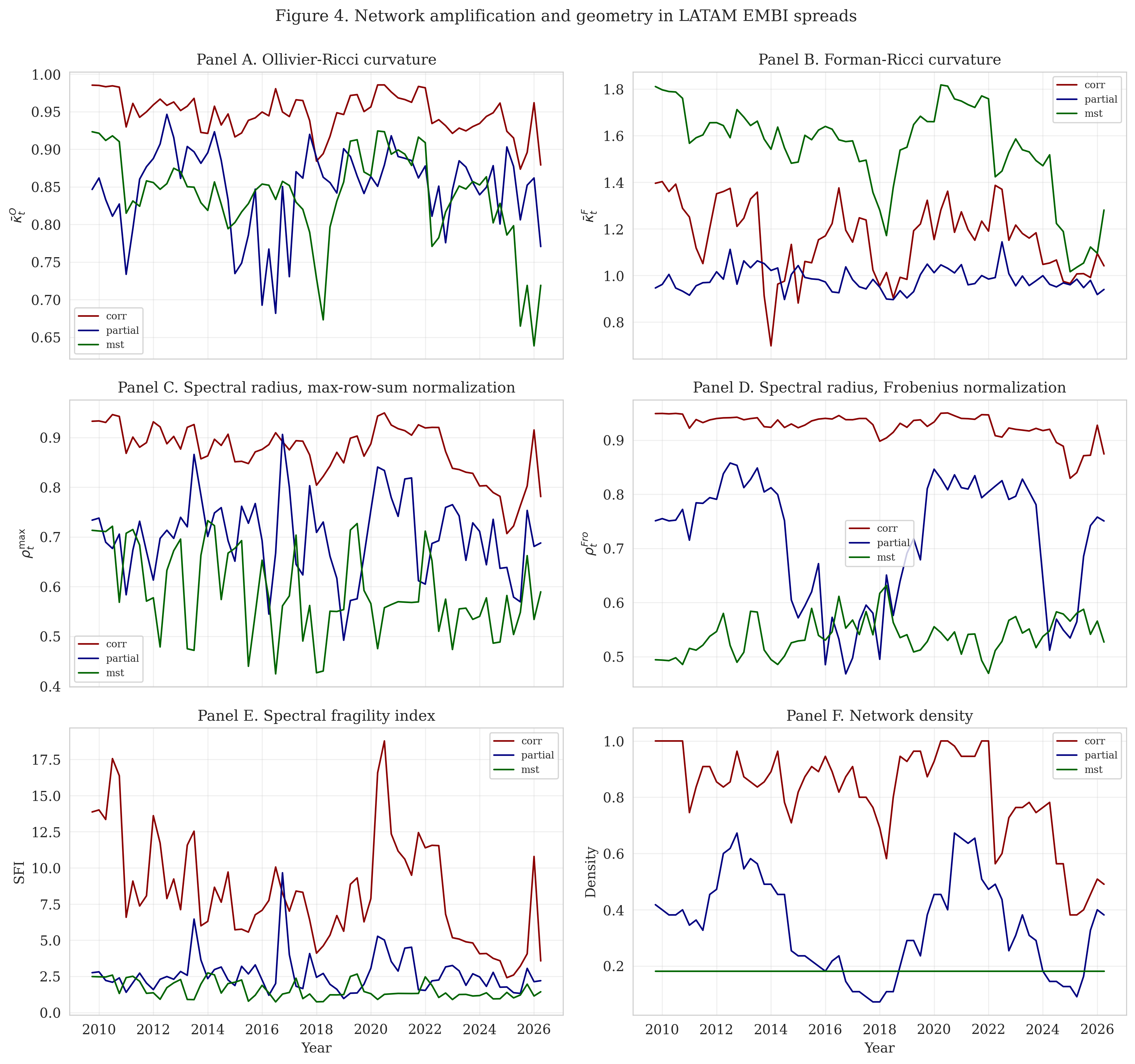}
\caption{Network amplification and geometry in LATAM EMBI spreads,
2009--2026. The figure reports curvature, normalized spectral radius,
Spectral Fragility Index, and density across correlation,
partial-correlation, and minimum spanning tree filters.}
\label{fig:geometry}
\end{figure}

Figure~\ref{fig:geometry} traces the dynamics of network amplification and
geometry. The most economically informative panels are the normalized
spectral radius, the Spectral Fragility Index, and network density.
Curvature measures are reported as auxiliary geometric indicators, while
spectral metrics are the main measures of amplification.

Table~\ref{tab:network_regime} compares network metrics in 24-month windows
containing at least one large avalanche ($S_t\geq 6$) with windows that do
not. The evidence on regime differences is heterogeneous across filters and
deserves a careful reading.

The correlation network displays the largest and most statistically
significant regime differences. Network density rises from $0.74$ in
non-large windows to $0.92$ in large-avalanche windows ($t=5.42$, $p<0.001$),
and the maximum-row-sum normalized spectral radius rises from $0.85$ to
$0.91$ ($t=6.26$, $p<0.001$). The corresponding maximum-row-sum SFI rises
from $6.23$ to $11.74$, and the Frobenius SFI from $11.86$ to $16.30$. The
HHI of node strength falls in large-avalanche windows ($t=-4.65$, $p<0.001$),
indicating that regional stress connectivity becomes both more amplifying and
more evenly distributed across sovereigns. The minimum spanning tree shows
mixed but smaller regime differences, reflecting the constraint of a fixed
edge count.

The partial-correlation network, by contrast, does \emph{not} display
significant regime differences in any of its spectral or density metrics. The
maximum-row-sum spectral radius is essentially flat ($0.871$ versus $0.855$,
$p=0.435$), the Frobenius spectral radius is mildly lower in large-avalanche
windows ($0.914$ versus $0.896$, $p=0.144$), and the corresponding SFI
measures show no statistically significant change. Density rises only
marginally ($0.69$ to $0.72$, $p=0.651$).

This contrast is informative rather than contradictory. Raw correlation
networks become denser and more spectrally amplifying around large
avalanches, but raw correlations are known to rise mechanically during
high-volatility episodes \citep{ForbesRigobon2002}. Partial correlations
filter the common linear factor that drives this mechanical effect. The fact
that partial-correlation networks do not exhibit additional spectral
amplification in large-avalanche windows therefore suggests that most of the
observed network-amplification signal in raw correlations is attributable to
common-factor co-movement rather than to genuine increases in conditional
regional propagation. \textbf{H5 is therefore partially confirmed: large
avalanche regimes are associated with denser and more amplifying network
states in the correlation filter, but this regime difference is not robust to
partial-correlation filtering.} The substantive interpretation is that
amplification in EMBI co-movement networks during large avalanches is
predominantly a common-factor phenomenon rather than a conditional-network
phenomenon.

\begin{table}[!htbp]
\centering
\footnotesize
\begin{threeparttable}
\caption{Network geometry around large stress months. Welch's two-sample $t$-test contrasts metric values in 24-month rolling windows containing a large avalanche ($S\geq 6$) versus all other windows.}
\label{tab:network_regime}
\begin{tabular}{llrrrrrrr}
\toprule
Filter & Metric & Non-large mean & Large mean & $\Delta$ & $t$ & $p$ & $n_L$ & $n_N$ \\
\midrule
corr & $\rho^{\max}$ & 0.847 & 0.911 & 0.065 & 6.259 & 0.000 & 33 & 34 \\
corr & $\rho^{\mathrm{Fro}}$ & 0.913 & 0.941 & 0.027 & 4.607 & 0.000 & 33 & 34 \\
corr & SFI$^{\max}$ & 6.229 & 11.743 & 5.515 & 6.304 & 0.000 & 33 & 34 \\
corr & SFI$^{\mathrm{Fro}}$ & 11.858 & 16.296 & 4.438 & 5.636 & 0.000 & 33 & 34 \\
corr & $\bar{\kappa}^O$ & 0.935 & 0.963 & 0.028 & 4.132 & 0.000 & 33 & 34 \\
corr & $\bar{\kappa}^F$ & 1.138 & 1.254 & 0.116 & 2.547 & 0.014 & 33 & 34 \\
corr & Density & 0.739 & 0.915 & 0.177 & 5.419 & 0.000 & 33 & 34 \\
corr & HHI strength & 0.106 & 0.096 & -0.010 & -4.645 & 0.000 & 33 & 34 \\
partial & $\rho^{\max}$ & 0.871 & 0.855 & -0.015 & -0.786 & 0.435 & 33 & 34 \\
partial & $\rho^{\mathrm{Fro}}$ & 0.914 & 0.896 & -0.018 & -1.480 & 0.144 & 33 & 34 \\
partial & SFI$^{\max}$ & 10.079 & 10.173 & 0.093 & 0.044 & 0.965 & 33 & 34 \\
partial & SFI$^{\mathrm{Fro}}$ & 12.696 & 11.159 & -1.537 & -1.244 & 0.219 & 33 & 34 \\
partial & $\bar{\kappa}^O$ & 0.923 & 0.909 & -0.014 & -0.878 & 0.383 & 33 & 34 \\
partial & $\bar{\kappa}^F$ & 1.262 & 1.230 & -0.032 & -0.533 & 0.596 & 33 & 34 \\
partial & Density & 0.693 & 0.715 & 0.023 & 0.454 & 0.651 & 33 & 34 \\
partial & HHI strength & 0.109 & 0.106 & -0.004 & -1.262 & 0.211 & 33 & 34 \\
mst & $\rho^{\max}$ & 0.562 & 0.614 & 0.052 & 2.832 & 0.006 & 33 & 34 \\
mst & $\rho^{\mathrm{Fro}}$ & 0.558 & 0.516 & -0.042 & -6.453 & 0.000 & 33 & 34 \\
mst & SFI$^{\max}$ & 1.336 & 1.721 & 0.385 & 2.988 & 0.004 & 33 & 34 \\
mst & SFI$^{\mathrm{Fro}}$ & 1.270 & 1.074 & -0.195 & -6.235 & 0.000 & 33 & 34 \\
mst & $\bar{\kappa}^O$ & 0.812 & 0.874 & 0.061 & 4.510 & 0.000 & 33 & 34 \\
mst & $\bar{\kappa}^F$ & 1.445 & 1.689 & 0.245 & 6.208 & 0.000 & 33 & 34 \\
mst & HHI strength & 0.135 & 0.119 & -0.016 & -5.930 & 0.000 & 33 & 34 \\
\bottomrule
\end{tabular}
\begin{tablenotes}\footnotesize
\item Filters refer to the network construction: $|\rho|>0.4$ (corr), $|\rho^{\mathrm{part}}|>0.4$ (partial), or the minimum-spanning-tree filter (mst).
\item Network density and edge counts in the MST are constant by construction, with 10 edges among 11 nodes, and are therefore omitted.
\item The correlation network exhibits the strongest large-vs-non-large discrimination: density, spectral radius, SFI, and curvature increase significantly during large-avalanche windows, while node-strength concentration falls.
\item Partial-correlation metrics do not display statistically significant regime differences, suggesting that the raw-correlation amplification signal is mainly driven by common-factor co-movement rather than conditional regional propagation.
\item MST metrics show statistically significant but mixed regime differences and should be interpreted as a sparse topological benchmark rather than as the main amplification evidence.
\end{tablenotes}
\end{threeparttable}
\end{table}

\subsection{Lead--lag network association (H6)}
\label{subsec:results_network_predictive}

Table~\ref{tab:network_predictive} reports lead--lag associations between
contemporaneous network metrics and the maximum avalanche size over the
subsequent three months,
$Y_t=\max_{u\in(t,t+3]} S_u$. The results are not interpreted as causal
estimates. They are intended to evaluate whether network states contain
early-warning information about near-future regional stress.

The empirical evidence on lead--lag predictability is weak. None of the
network metrics in any of the three filters yields a Pearson correlation that
is statistically significant at conventional levels. The largest absolute
correlation is observed for the maximum-row-sum spectral radius of the
correlation network, $r=+0.153$ ($p=0.217$), which suggests a positive but
imprecisely estimated association. The corresponding metrics in the
partial-correlation network are smaller in absolute value and not
statistically distinguishable from zero, with the Frobenius spectral radius
giving $r=-0.097$ ($p=0.436$) and the Frobenius SFI giving $r=-0.073$
($p=0.555$). The minimum spanning tree shows similarly weak associations.

The point estimates are also small in economic terms. With $n=67$ rolling
network snapshots, even the strongest observed correlation explains less than
$3$ percent of the variance in three-month-ahead maximum avalanche size.
H6 is therefore not supported by the data: contemporaneous network
metrics do not provide statistically significant predictive content for
near-future regional stress at the three-month horizon in this sample.

\begin{table}[!htbp]
\centering
\footnotesize
\begin{threeparttable}
\caption{Lead--lag association between current network geometry and the maximum avalanche size over the next three months.}
\label{tab:network_predictive}
\begin{tabular}{llrrr}
\toprule
Filter & Metric & Pearson $r$ & $p$ & $n$ \\
\midrule
corr & $\rho^{\max}$ & +0.153 & 0.217 & 67 \\
corr & $\rho^{\mathrm{Fro}}$ & +0.094 & 0.448 & 67 \\
corr & SFI$^{\max}$ & +0.080 & 0.518 & 67 \\
corr & SFI$^{\mathrm{Fro}}$ & +0.077 & 0.537 & 67 \\
corr & $\bar{\kappa}^O$ & +0.027 & 0.826 & 67 \\
corr & Density & +0.117 & 0.345 & 67 \\
corr & HHI strength & -0.128 & 0.304 & 67 \\
partial & $\rho^{\max}$ & -0.117 & 0.347 & 67 \\
partial & $\rho^{\mathrm{Fro}}$ & -0.097 & 0.436 & 67 \\
partial & SFI$^{\max}$ & +0.021 & 0.867 & 67 \\
partial & SFI$^{\mathrm{Fro}}$ & -0.073 & 0.555 & 67 \\
partial & $\bar{\kappa}^O$ & -0.040 & 0.749 & 67 \\
partial & Density & -0.006 & 0.960 & 67 \\
partial & HHI strength & -0.038 & 0.758 & 67 \\
mst & $\rho^{\max}$ & +0.121 & 0.329 & 67 \\
mst & $\rho^{\mathrm{Fro}}$ & -0.059 & 0.638 & 67 \\
mst & SFI$^{\max}$ & +0.132 & 0.287 & 67 \\
mst & SFI$^{\mathrm{Fro}}$ & -0.053 & 0.672 & 67 \\
mst & $\bar{\kappa}^O$ & +0.063 & 0.613 & 67 \\
mst & HHI strength & -0.060 & 0.631 & 67 \\
\bottomrule
\end{tabular}
\begin{tablenotes}\footnotesize
\item Each row reports the Pearson correlation between a network metric at month $t$ and $\max_{u\in(t,t+3]} S_u$.
\item No network metric displays a statistically significant three-month-ahead association with future maximum avalanche size.
\item The largest absolute correlation is modest, $|r|=0.153$, and all reported $p$-values exceed conventional significance levels.
\item MST metrics should be interpreted cautiously because the minimum-spanning-tree filter imposes a fixed-edge topology.
\end{tablenotes}
\end{threeparttable}
\end{table}

This null result deserves direct discussion rather than being explained
away. Three considerations are relevant. First, the predictive horizon and
sample size are modest: with monthly rolling windows and a forecast horizon
of three months, the effective number of independent observations available
for inference is small, and statistical power is limited. Second, the
finding is consistent with the regime-comparison evidence in
Section~\ref{subsec:results_network_regime}: most of the network signal
around large avalanches reflects common-factor co-movement, which is
contemporaneous with stress rather than a leading indicator of it. Third, the
result does not preclude predictive content at different horizons, with
different network filters, or in specifications that combine network
information with global risk variables. A richer multivariate predictive
model is left to future work.

The conservative reading is that the network evidence in this paper
identifies a contemporaneous regime structure (Section~\ref{subsec:results_network_regime}) but does not establish a robust
early-warning signal at the three-month horizon. Operationally, this means
that the avalanche-size process $S_t$ and the placebo-validated
synchronization evidence remain the primary surveillance objects of the
paper, while network amplification metrics are best interpreted as
descriptive features of stress regimes rather than as predictive instruments.

\subsection{Summary of findings}
\label{subsec:results_summary}

Table~\ref{tab:hypothesis_summary} summarizes the six empirical hypotheses
in the order in which they were stated in Section~\ref{sec:theory}. The
evidence supports the presence of heavy-tailed finite-size avalanches (H1),
non-random synchronization (H2), threshold robustness over the informative
range (H3), and heterogeneous country participation (H4). The evidence on
network amplification during large avalanches (H5) is partially supportive,
with the correlation filter showing strong regime differences but the
partial-correlation filter showing none, suggesting that the regime
difference reflects common-factor co-movement rather than conditional
network amplification. The evidence on predictive lead--lag network
association (H6) is not supportive at the three-month horizon in this
sample. The strongest results of the paper are therefore the
synchronization placebo and the threshold-robust existence of large regional
avalanches.

\begin{table}[!htbp]
\centering
\footnotesize
\begin{threeparttable}
\caption{Summary of hypotheses, tests, and outcomes.}
\label{tab:hypothesis_summary}
\begin{tabular}{lp{0.27\linewidth}p{0.30\linewidth}p{0.27\linewidth}}
\toprule
\# & Hypothesis & Test / evidence & Outcome \\
\midrule

H1 & Avalanche-size distribution is heavy-tailed and lies in a finite-size criticality regime. 
& Clauset MLE + Vuong test; Table~\ref{tab:tail_diagnostics}; Fig.~2. 
& Confirmed in conservative form; $\hat{\alpha}=1.77$, $x_{\min}=1$, and the lognormal alternative is not decisively preferred. \\

H2 & Cross-country sovereign-stress synchronization is non-random. 
& Country-level reshuffling placebo, $B=1{,}000$; Table~\ref{tab:placebo}. 
& Confirmed decisively; observed $S_{\max}=11$ is far above the placebo distribution, with $p<0.001$ for all synchronization statistics. \\

H3 & Avalanche evidence is robust to the choice of stress threshold $\sigma$. 
& Threshold sensitivity over $\sigma\in[1.25,2.50]$; Table~\ref{tab:threshold_robustness}; Fig.~4. 
& Confirmed over the informative range; $\hat{\alpha}\in[1.70,1.83]$ for $\sigma\leq2.0$, and $S_{\max}=11$ for all tested thresholds. \\

H4 & Country-level participation in stress avalanches is heterogeneous and not mechanically ordered by average spread levels. 
& Country event counts; Table~\ref{tab:country_thresholds}; Fig.~5--6. 
& Confirmed; participation ranges from 7 to 18 events, with Brazil, Mexico, Peru, Colombia, and Panama among the most recurrent participants. \\

H5 & Large-avalanche windows are associated with more amplifying contemporaneous network states. 
& Regime comparison between large-avalanche and non-large-avalanche windows; Table~\ref{tab:network_regime}; Fig.~8. 
& Partially confirmed; raw correlation networks become denser and more spectrally amplifying, but the result is not robust to partial-correlation filtering. \\

H6 & Network amplification contains predictive lead--lag information for near-future regional stress. 
& Three-month-ahead Pearson correlations and OLS associations; Table~\ref{tab:network_predictive}. 
& Not supported; no network metric is statistically significant at the three-month horizon in any filter. \\

\bottomrule
\end{tabular}
\begin{tablenotes}\footnotesize
\item All tests are based on monthly EMBI Global Diversified stripped spreads for 11 LATAM countries over 2007:M10--2026:M4.
\item The heavy-tail evidence is interpreted conservatively as finite-size criticality evidence, not as proof of a unique asymptotic power law.
\item The mid-credibility result is treated as part of H4's substantive interpretation rather than as a separate formal hypothesis.
\end{tablenotes}
\end{threeparttable}
\end{table}

Overall, the empirical results indicate that Latin American sovereign credit
markets exhibit finite-size avalanche dynamics. The system is not merely a
collection of independent national spread shocks. Large regional stress
episodes are synchronized beyond country-level hit frequencies, appear in
historically meaningful periods, and are associated with denser correlation
network states during stress regimes. The evidence is therefore consistent
with a finite-size criticality interpretation of LATAM sovereign stress in
its avalanche-size and synchronization dimensions, while the
network-amplification dimension is more conditional: it is detectable in raw
co-movement but not robust to partial-correlation filtering, and it does not
deliver statistically significant lead--lag predictive content at the
three-month horizon. The paper therefore claims a heavy-tailed, synchronized,
and historically coherent regional stress process, while remaining cautious
about the stronger claims of conditional network amplification and
network-based early warning.

\section{Discussion}
\label{sec:discussion}

The empirical results position Latin American sovereign credit markets in a
finite-size heavy-tail regime. The evidence is not strong enough to claim
that EMBI spreads obey a unique asymptotic power law, nor that Latin
American sovereign risk is a pure self-organized critical system. However,
the results are clearly inconsistent with a thin-tailed,
independent-country interpretation of sovereign stress. Large regional
avalanches occur substantially more frequently than would be expected
under country-level stress frequencies alone, and these avalanches occur in
contemporaneous network environments that are denser and more spectrally
integrated. The relevant interpretation is therefore one of finite-size
criticality: a bounded regional system in which stress is episodic,
synchronized, heavy-tailed, and embedded in a time-varying co-movement
structure that amplifies common shocks. The paper is also explicit about
what the network evidence does not establish, in particular the absence of
significant lead--lag predictive content at the three-month horizon and
the absence of regime differences once partial-correlation filtering
removes common-factor co-movement.

\subsection{Implications for surveillance}

The first implication concerns sovereign-risk surveillance. Standard
monitoring frameworks focus on country-level fundamentals: debt ratios,
primary balances, reserves, current-account positions, inflation, rollover
needs, and political risk. These variables remain essential, but the
evidence in this paper suggests that they should be complemented by
regional synchronization indicators. The avalanche-size process $S_t$
provides a simple, computationally inexpensive measure of how many
sovereigns are simultaneously experiencing large spread innovations, and
its empirical distribution and historical alignment with global systemic
events make it a meaningful regional stress-state variable.

The role of network metrics in this surveillance framework requires more
careful articulation than the paper initially anticipated. The
regime-comparison evidence in Table~\ref{tab:network_regime} shows that
the raw correlation network discriminates large-avalanche from
non-large-avalanche windows with high statistical significance: density,
the maximum-row-sum spectral radius, the Spectral Fragility Index, and
the inverse of node-strength concentration all move sharply when the
regional system enters a stress regime. The partial-correlation network,
by contrast, does not discriminate regimes in any of its spectral or
density metrics. The substantive interpretation of this asymmetry is that
most of the network amplification observed during large avalanches
reflects common-factor co-movement of the type emphasized by
\citet{ForbesRigobon2002}, rather than additional conditional regional
propagation. Once the common linear factor is filtered out, the network
no longer becomes more amplifying around stress episodes.

This finding does not eliminate the operational usefulness of the SFI; it
clarifies what kind of usefulness it has. The SFI on the raw correlation
network behaves as a contemporaneous descriptor of regional stress
regimes, in the same way that realized volatility, implied volatility, or
the breadth of asset-class drawdowns are contemporaneous descriptors of
financial conditions. The lead--lag analysis in
Table~\ref{tab:network_predictive} reinforces this contemporaneous
interpretation: no network metric, in any of the three filters, exhibits
a statistically significant correlation with three-month-ahead maximum
avalanche size. The largest absolute correlation in the table is
$r=+0.153$ for the maximum-row-sum spectral radius of the correlation
network, with $p=0.217$. This null result deserves direct acknowledgment
rather than rationalization: the network metrics in this sample are
informative about contemporaneous stress regimes, but they do not deliver
robust early-warning signals at the three-month horizon.

In practice, this means that monthly SFI nowcasts can be computed at low
cost from rolling EMBI spread changes and used as an additional layer of
contemporaneous regional stress monitoring by central banks, multilateral
institutions, and sovereign-risk teams. They should not be interpreted as
forecasts of crisis. Their role is closer to that of a thermometer than
that of a barometer: they describe the current temperature of the
regional stress system rather than predict where it is going. For
genuine early-warning content, future work would need to combine network
metrics with other variables---global risk appetite indices, U.S.\
monetary-policy indicators, commodity prices, and country-level
fundamentals---in a richer multivariate framework that the present paper
does not attempt.

\subsection{Regional political economy of sovereign fragility}

The second implication concerns the political economy of regional
fragility. A conventional weak-link view would suggest that the most
fragile countries are the highest-spread sovereigns. Under that
interpretation, Argentina and Ecuador should dominate the stress-event
process. The data do not support that view. These countries display very
high spread levels and extreme regime-defining episodes, but they do not
generate the largest number of threshold crossings. Their stress is
episodic and discontinuous: defaults, currency collapses, and
restructuring episodes can push spreads into high absorbed states, after
which additional monthly threshold crossings become less frequent.

By contrast, countries such as Brazil, Mexico, Peru, Colombia, and Panama
participate more frequently in regional stress events. These sovereigns
are not necessarily the highest-risk economies in average spread terms,
but they are more liquid, more investable, and more central in global
emerging-market portfolios. Their spreads respond repeatedly to changes
in global risk appetite, monetary tightening, benchmark rebalancing, and
regional investor coordination. The implication is that Latin American
sovereign fragility is not only the fragility of its weakest fiscal
links. It is also the fragility of its capital-market integration as a
system. The countries that matter most for regional avalanches may be
those that sit at the intersection of credibility, liquidity, and
exposure to global portfolio adjustment.

This interpretation is reinforced by the partial-correlation regime
result. If the network amplification observed during large avalanches
were driven primarily by structural regional propagation, conditional
co-movement should rise during stress episodes. The fact that it does
not---while raw co-movement does---is consistent with the view that the
mid-credibility, highly-integrated sovereigns participate in regional
stress because they jointly absorb global shocks rather than because they
transmit shocks to one another in conditional terms. Latin American
sovereign credit markets behave, during stress regimes, more as a single
asset class facing a common factor than as a network of bilateral
propagation channels.

This distinction is important for policy interpretation. A surveillance
framework focused only on fiscal weakness may identify countries with
high default risk but miss the countries that most strongly synchronize
regional stress through their integration into global emerging-market
portfolios. Conversely, a network-based framework that emphasizes
conditional regional propagation may overstate the role of bilateral
transmission relative to common-factor exposure. For regional financial
stability, the relevant question is therefore not only which country is
fiscally weakest, but which sovereigns are positioned to convert global
risk shocks into regional co-movement, and what fraction of regional
synchronization is structural versus common-factor in nature.

\subsection{Criticality as a system-level property}

The third implication is theoretical. The results suggest that
criticality in sovereign credit markets is a system-level phenomenon
rather than a smooth aggregation of country-level properties. Additional
country-level diagnostics show that the association between
country-specific recurrence exponents and country-level curvature is weak
and statistically insignificant. This finding is not a failure of the
framework. It is consistent with the logic of sandpile-type systems,
where critical behavior is not a property of isolated sites but an
emergent feature of the interaction structure.

In the present context, no single country needs to display a stable
curvature--exponent relationship for the regional system to generate
heavy-tailed avalanches. What matters is the configuration of thresholds,
market integration, common shocks, and conditional co-movement. Large
avalanches emerge when enough countries are simultaneously close to their
stress thresholds and are jointly exposed to a sufficiently strong
common factor. This is why the paper emphasizes system-level
objects---$S_t$, density, spectral radius, SFI, and the placebo
synchronization test---rather than relying on country-level tail
estimates alone.

This interpretation also clarifies the role of curvature. The empirical
Ollivier--Ricci and Forman--Ricci curvature measures are useful as
geometric summaries of network integration, but they should not be
treated as the central fragility indicator in this application. In dense
sovereign-spread networks, positive curvature may reflect redundant
co-movement rather than bottleneck fragility. The main operational
measures of amplification are therefore spectral: the normalized spectral
radius and the corresponding Spectral Fragility Index. These measures
connect more directly to the propagation dynamics and to the stability
boundary of linearized network systems, even if their predictive content
in this sample is contemporaneous rather than leading.

\subsection{Limitations and extensions}

Several limitations qualify the interpretation. First, EMBI spreads are
secondary-market price indicators. They capture sovereign credit risk as
priced by investors, but they do not directly observe bond flows, foreign
holdings, dealer balance sheets, issuance quantities, or primary-market
funding costs. The avalanche process identified here is therefore an
asset-price stress process, not a full balance-sheet contagion mechanism.

Second, the network is inferred from rolling co-movement in monthly
spread changes. This is appropriate for a long regional panel, but it
cannot fully separate structural transmission from common shocks,
portfolio rebalancing, or global risk-appetite shifts. The
partial-correlation network reduces this problem and, as discussed
above, the comparison between the two filters is itself informative
about the common-factor versus conditional-network decomposition. Future
work could combine EMBI spreads with fund-flow data, mutual-fund
holdings, sovereign issuance calendars, or high-frequency CDS data to
better identify transmission channels and to test whether the
common-factor interpretation continues to dominate at higher frequencies.

Third, the analysis is finite-sample by construction. The system contains
only eleven sovereigns, the monthly sample covers $223$ periods, and the
number of avalanche events above the estimated power-law cutoff is
moderate. This is adequate for documenting finite-size avalanche behavior
and non-random synchronization, but it limits the ability to distinguish
cleanly between power-law, lognormal, and other heavy-tailed
alternatives, and it limits the power of the lead--lag predictive
analysis. For this reason, the paper does not claim to prove an
asymptotic SOC law, and it does not interpret the null lead--lag result
as definitive evidence against predictive content; both conclusions are
constrained by sample size.

Fourth, the MST filter is intentionally sparse and has a fixed number of
edges. This makes it useful as a topological benchmark but limits its
ability to capture high-density stress regimes. Richer filtering methods,
such as planar maximally filtered graphs \citep{Tumminello2005},
graphical lasso networks, dynamic factor-adjusted networks, or
time-varying parameter VAR connectedness models, could provide additional
robustness. Similarly, the 24-month rolling window used here balances
precision and responsiveness, but future work could validate the results
using weekly or daily EMBI/CDS data, where the contemporaneous-versus-
predictive distinction can be sharpened.

\subsection{What the paper does and does not establish}

It is useful to state the interpretive boundaries of the paper directly.
The paper establishes that LATAM sovereign credit markets exhibit
heavy-tailed finite-size avalanche behavior, threshold-robust large
regional stress events, and synchronization that is far stronger than
what would be achievable under independent country-level timing. It
establishes that participation in these avalanches is dominated by
mid-credibility, highly integrated sovereigns rather than by the
highest-spread economies. It establishes that the contemporaneous
co-movement network becomes denser and more spectrally amplifying
during stress regimes when measured by raw correlations.

The paper does not establish that EMBI spreads obey a unique asymptotic
power law, that the avalanche process is governed by a pure SOC
mechanism, or that conditional regional propagation increases
significantly during stress regimes once common-factor co-movement is
filtered out. It does not establish a statistically significant lead--lag
predictive relationship between network metrics and regional stress at
the three-month horizon. These boundaries are not weaknesses of the
analysis. They define the empirical content of the paper precisely: a
disciplined, finite-sample documentation of heavy-tailed and synchronized
sovereign stress in a region for which such evidence has not previously
been assembled.

Overall, the evidence supports a disciplined interpretation: Latin
American sovereign credit markets display finite-size criticality
signatures, but not a uniquely identified pure SOC law. The system
generates large synchronized stress avalanches, these avalanches are
historically meaningful and statistically non-random, and their
occurrence coincides with denser regional co-movement networks. This
combination of results suggests that sovereign-risk surveillance should
move beyond country-by-country spread monitoring toward a regional view
of stress synchronization and contemporaneous network state.

\section{Conclusion}
\label{sec:conclusion}

This paper studies whether Latin American sovereign credit markets exhibit
empirical signatures of finite-size criticality. Using monthly J.P.~Morgan
EMBI Global Diversified stripped spreads for eleven Latin American
economies from October 2007 to April 2026, the paper constructs
country-level stress events from threshold exceedances in monthly
log-spread changes and defines regional sovereign-stress avalanches as
the number of countries simultaneously in stress. The analysis combines
heavy-tail diagnostics, threshold robustness checks, reshuffling
placebos, rolling co-movement networks, spectral amplification measures,
and lead--lag network associations.

The evidence supports a conservative but economically meaningful
conclusion: Latin American sovereign spreads do not behave as a
collection of independent country-level shocks. Instead, regional stress
appears as a synchronized and heavy-tailed process embedded in a
time-varying co-movement environment. Avalanche sizes exhibit a
heavy-tailed distribution with an estimated exponent in the finite-size
criticality range. Absolute spread changes and inter-event times also
fall within a heavy-tail boundary regime in the sense of
\citet{StumpfPorter2012}: power-law and lognormal specifications cannot
be sharply separated in finite samples, but thin-tailed alternatives are
inconsistent with the observed patterns. The results therefore do not
prove a pure asymptotic self-organized critical law. They provide
evidence that LATAM sovereign credit markets display finite-size
avalanche dynamics consistent with critical propagation.

The strongest empirical result is the synchronization placebo. Preserving
each country's stress frequency but randomly reshuffling stress timing
within country generates substantially smaller avalanches than those
observed in the data, with $p<0.001$ across all four synchronization
statistics. This indicates that large regional sovereign-stress episodes
are not a mechanical consequence of having volatile countries in the
sample. They are temporally organized events. The largest avalanches
coincide with historically meaningful episodes such as the Lehman
collapse, the post-Lehman emerging-market rout, and the COVID-19 shock,
reinforcing the interpretation of $S_t$ as a regional stress-state
variable.

The network results add a contemporaneous regime dimension that requires
careful articulation. Large avalanche regimes are associated with denser
and more spectrally amplifying co-movement networks when measured by raw
correlations. However, this regime difference is not robust to
partial-correlation filtering: once the common linear factor is removed,
the spectral and density metrics no longer discriminate large-avalanche
from non-large-avalanche windows. The substantive interpretation is that
the observed network amplification reflects common-factor co-movement
rather than additional conditional regional propagation, consistent with
the well-known sensitivity of raw correlations to high-volatility
episodes \citep{ForbesRigobon2002}. Lead--lag associations between
contemporaneous network metrics and three-month-ahead maximum avalanche
size are not statistically significant in any of the three network
filters, with the largest absolute Pearson correlation reaching only
$r=+0.153$ ($p=0.217$). The Spectral Fragility Index and related metrics
are therefore best interpreted as descriptors of contemporaneous regional
stress regimes rather than as early-warning instruments at the
three-month horizon.

The results also revise the political economy interpretation of
sovereign fragility in Latin America. Regional stress is not driven only
by the highest-spread or most fiscally distressed sovereigns. Countries
such as Brazil, Mexico, Peru, Colombia, and Panama participate frequently
in stress events because they are liquid, investable, and deeply
connected to global emerging-market portfolio dynamics. Argentina and
Ecuador remain important high-risk sovereigns, but their stress episodes
are more discontinuous and regime-specific. The relevant source of
regional fragility is therefore not only weak fiscal fundamentals at the
periphery; it is also the systemic integration of mid-credibility
sovereigns into a common market for emerging-market credit risk. The
asymmetry between the correlation and partial-correlation regime
results reinforces this reading: regional fragility appears as the joint
exposure of integrated sovereigns to a common risk factor, rather than
as the product of dense bilateral propagation channels.

The contribution of the paper is methodological and empirical.
Methodologically, it provides a framework for measuring sovereign-stress
avalanches, testing whether synchronization exceeds country-level stress
frequencies, and characterizing the contemporaneous network environment
of regional stress regimes. The disciplined separation between
correlation and partial-correlation filters, and between contemporaneous
regime comparisons and lead--lag predictive associations, is itself a
methodological contribution: it shows how to extract honest empirical
content from network metrics in a setting where common-factor exposure is
a dominant force. Empirically, it shows that LATAM EMBI spreads display
finite-size avalanche behavior, threshold-robust large synchronized
events, and contemporaneous network states that become more
correlation-dense around systemic episodes. This complements standard
sovereign-spread models based on fundamentals, liquidity, and global
risk appetite by adding a regional dynamic layer: the same country-level
fundamentals may generate different systemic outcomes depending on the
contemporaneous state of the sovereign co-movement network.

The findings are relevant for surveillance. Central banks, multilateral
institutions, and sovereign-risk analysts typically monitor country-level
spreads, fiscal indicators, external balances, and rollover needs. The
results suggest that these indicators should be complemented by
system-level measures: the avalanche-size process, the frequency of
large synchronized events, and contemporaneous network descriptors such
as correlation density and the Spectral Fragility Index. These measures
are computationally simple, can be updated at monthly frequency, and
provide information about the current regional amplification state of
the sovereign credit system. They should not be presented to
policymakers as predictive instruments, but as nowcasting tools that
sharpen the contemporaneous picture of regional stress.

Several extensions remain open. Future work could apply the framework
to weekly or daily EMBI/CDS data, where the timing of stress propagation
can be measured more precisely and where the contemporaneous-versus-
predictive distinction can be tested at horizons short enough to make
network amplification potentially leading rather than coincident. The
approach could also be extended to other emerging-market regions to
assess whether finite-size sovereign avalanches are a Latin American
phenomenon or a more general feature of emerging-market debt. A further
extension would be to combine EMBI data with fund-flow, holdings,
issuance, foreign-exchange, commodity-price, and macro-fiscal variables
in order to move from descriptive contemporaneous association toward
structural identification of the propagation mechanism, and to test
whether richer multivariate predictive specifications can recover
significant lead--lag content that the parsimonious bivariate analysis
in this paper does not detect. Finally, the framework could be used for
counterfactual policy simulations, evaluating whether shocks to specific
sovereigns, global risk factors, or network links produce materially
different regional avalanche distributions.

In sum, the paper does not claim that Latin American sovereign credit
markets are governed by a pure self-organized critical law. It shows
that they exhibit several of the empirical ingredients of finite-size
criticality---heavy-tailed stress events, non-random synchronization,
threshold-robust regional avalanches, and a contemporaneous network
environment that becomes denser and more spectrally integrated during
stress---while also showing that the network amplification signal is
largely common-factor in nature and does not deliver robust early-warning
content at the three-month horizon. For sovereign-risk surveillance, the
implication is straightforward: regional fragility should be monitored
not only through country fundamentals, but also through the dynamic
geometry of sovereign credit co-movement, with appropriate attention to
the distinction between contemporaneous regime descriptors and forward-
looking predictive instruments.

\bibliography{references}

\end{document}